# Cabin Layout, Seat Density, and Passenger Segmentation in Air Transport: Implications for Prices, Ancillary Revenues, and Efficiency

Alessandro V. M. Oliveira⁺
Aeronautical Technology Institute - ITA
Moisés D. Vassallo
Federal University of Itajubá - UNIFEI
⁺Corresponding author. Praça Marechal Eduardo Gomes, 50. São José dos Campos, SP - Brazil.
E-mail address: alessandro@ita.br .

*Abstract*: This study investigates how the layout and density of seats in aircraft cabins influence the pricing of airline tickets on domestic flights. The analysis is based on microdata from boarding passes linked to face-to-face interviews with passengers, allowing us to relate the price paid to the location on the aircraft seat map, as well as market characteristics and flight operations. Econometric models were estimated using the Post-Double-Selection LASSO (PDS-LASSO) procedure, which selects numerous controls for unobservable factors linked to commercial and operational aspects, thus enabling better identification of the effect of variables such as advance purchase, reason for travel, fuel price, market structure, and load factor, among others. The results suggest that a higher density of seat rows is associated with lower prices, reflecting economies of scale with the increase in aircraft size and gains in operational efficiency. An unexpected result was also obtained: in situations where there was no seat selection fee, passengers with more expensive tickets were often allocated middle seats due to purchasing at short notice, when the side alternatives were no longer available. This behavior helps explain the economic logic behind one of the main ancillary revenues of airlines. In addition to quantitative analysis, the study incorporates an exploratory approach to innovative cabin concepts and their possible effects on density and comfort on board.

*Keywords*: seat density, cabin layout, airfare pricing, density economies, operational efficiency, revenue management, LASSO modeling, high-dimensional inference.

## I. INTRODUCTION

The internal configuration of commercial aircraft, especially the layout and density of seats, plays a decisive role in the cost structure and perceived value of air services. The decision on how many rows to install and how much space to offer between seats involves a balance between operational efficiency and perceived passenger comfort, elements that directly influence the formation of operating costs, ticket prices, and the market segmentation of airlines.

In the context of Brazilian domestic aviation, dominated by the operation of narrowbody aircraft in economy class, companies have adopted product differentiation strategies based on cabin layout since the last decade. This movement included the offer of premium economy versions characterized mainly by additional space between seats, in addition to increased charges for advance seat selection. This movement differs from the international segment, where segmentation and class distinction practices were already more consolidated from previous periods. These transformations occurred simultaneously with the spread of cost-efficiency-oriented business models, in which increased seat density is seen as an effective mechanism for reducing cost per seat-kilometer (CASK). However, the quantitative effects of this density on the prices actually paid by passengers are still poorly understood, especially in domestic markets with concentrated competitive structures and highly heterogeneous traveler profiles.

This study analyzes the relationship between cabin layout, seat density, and airfare prices in the Brazilian domestic market. The analysis uses an unprecedented database constructed by linking face-to-face interviews with passengers and microdata from boarding passes collected in 2014. This combination allows us to identify the amount actually paid by each traveler and relate it accurately to the characteristics of the flight, the seat, the internal configuration of the aircraft, and the operator. The study also develops a qualitative discussion on how cabin layout influences comfort, perceived value, and commercial strategies of airlines. This discussion addresses design innovations proposed in recent years by manufacturers and developers, including the Morph Seating, Cozy Suite, Air Lair, StepSeat, FlexSeat, Zephyr Seat, Cabin Hexagon, Checkerboard, and Skyrider projects. Each of these proposals involves structural changes in the use of internal space and may affect operating costs, cabin density, ergonomics, and possible variations in passengers' willingness to pay. Thus, the study combines econometric analysis and conceptual reflection to examine the role of aircraft physical configuration in price formation.

The database used contains detailed information on passenger purchasing profiles. By linking these elements to the seat occupied by the traveler, it is possible to observe whether structural differences in the cabin and passenger location are statistically associated with the price paid. To explore this relationship, the Post-Double-Selection LASSO (PDS-LASSO) econometric method was used. This method performs variable selection in high-dimensional environments and reduces some of the omitted variable bias by identifying subsets of controls related to both price and variables of interest. The initial set of controls included effects of survey date, flight time, origin and destination airports, and individual passenger characteristics. These groups were subjected to a selection procedure that extracts statistically relevant controls for each specification. This approach contributes to improving the accuracy of estimates, although it maintains limitations inherent to regularization methods and the structure of the data used.

The paper is structured as follows. After this introduction, Section II presents a review of the literature on cabin layout, comfort, and market effects. Section III discusses the case study, including the historical evolution of configurations in Brazil, a description of innovative concepts in the industry, and an exploratory visualization of a high-density cabin layout. Next, the conceptual framework that supports empirical modeling is discussed in Section IV, followed by the presentation of data in Section V and methodological and econometric procedures in Section VI. Section VII presents the results of the estimates and, finally, Section VIII summarizes the conclusions of the study.

## II. CABIN LAYOUT, COMFORT, AND MARKET EFFECTS

The literature on the effects of cabin interior configuration in relation to aircraft seating is not abundant. In general, it can be divided into the following topics: I. level of comfort perceived by passengers; II. innovations and commercial practices of airlines; and III. willingness to pay for comfort and differentiation of airline products. Below is a review of this literature.

### II.1. LEVEL OF COMFORT PERCEIVED BY PASSENGERS

Kremser et al. (2012), Miller, Lapp & Parkinson (2019), and Anjani et al. (2020) investigate the relationship between seat pitch[1] and passenger-perceived comfort on board. Miller, Lapp & Parkinson (2019) include passenger load factor and demographics in their analysis of this relationship. Anjani et al. (2020) contribute by analyzing other influencing factors, such as spatial experience and anthropometric measurements. These authors conducted a survey of 294 participants who experienced economy class seats on a Boeing 737 with varying seat pitches between 28 and 34 inches. In the survey, anthropometric measurements were collected from the participants. The authors asked respondents questions about their perceived levels of comfort and discomfort. To define "spatial experience," the researchers presented statements to be evaluated by the respondents, such as "*I feel restricted by the distance between the rows of seats*" and "*I feel stressed because of the distance between the rows of seats*," among others. The study found a significant relationship between seat pitch and passenger comfort and discomfort measures. In addition, evidence was obtained that the average rating given by respondents to the discomfort level of each pitch in relation to the middle seat was systematically higher than that of the window and aisle seats. However, they attested that seat pitch apparently affects passenger (dis)comfort more than its location in the cabin. Finally, they obtained evidence that anthropometric measurements significantly affect (dis)comfort in smaller seat pitch sizes.

In Brazil, the topic of passenger comfort and its relationship with the anthropometric profile of the Brazilian population, as well as aircraft seat designs, was addressed in a series of studies between the late 2000s and early 2010s. Among these studies, Silva & Monteiro (2009), Souza (2010), Rossi (2011), Greghi (2012), and Silva Filho, Andrade & Ciaccia (2012) stand out.

---

[1] The seat pitch is the distance between any point on a seat and the same point on the seat in the same position in the row immediately in front or behind, measured in centimeters or inches. In general, economy class seat pitches range from 28 to 34 inches, or 71 to 86 cm (rounded, since one inch equals 2.54 cm).

## II.2. INNOVATIONS AND COMMERCIAL PRACTICES OF AIRLINES

Rothkopf & Wald (2011) study the introduction of innovations in the service provided by airlines, such as in-flight entertainment with individual LED screens in economy class seats, web check-in, airport service kiosks, and the use of smartphones on board. Some of the innovations related to the internal layout of the aircraft discussed are the policy of blocking the middle seat in business class on short-haul flights in narrow-body aircraft and the use of a "hybrid cabin," where there is a distinction in services within economy class, such as the creation of premium economy class with larger seat pitch. The authors analyze a sample of thirty airlines and detect patterns of innovations used by them within several identified categories of innovation. The authors suggest that there are different innovation priorities for the different business models of existing airlines.

Most studies in the literature related to the economic issue of seat design in commercial aviation focus on the issue of airline seat pricing (fares) and its consequences. Below, we present some of these studies.

Kyparisis & Koulamas (2018) study the price and ideal allocation of seats for an airline revenue management problem, considering the existence of two cabins, in which there is a flexible partition between business and economy cabins. The authors identify the optimal cabin division and optimal fares for both cabins with a general mathematical formulation based on a multiplicative price-demand function. They consider three different random distributions of travel demand. Additionally, they present evidence that the optimal partitioning and optimal prices are not sensitive to random demand distributions. The study also addresses the effect of a capacity constraint in the business cabin and concludes that this constraint would have the effect of raising business class fares and generating a drop in total revenue.

Mumbower, Garrow & Newman (2015) study purchases of seat selection services in premium economy class and their implications for airlines' optimal pricing strategies. The authors use a database of prices and online seat map displays collected from the website of the US airline JetBlue Airways. The results of the study indicate that multiple factors influence online purchasing behavior, including the seat selection fee, advance purchase, the number of passengers traveling together, and aircraft utilization factors—observed online as the flight seat map was displayed. The study provides evidence that customers are between 2 and 3.3 times more likely to purchase premium economy class seats—configured with extra legroom and early boarding privileges—when there are no window or aisle seats left on the entire aircraft that can be purchased for free. The authors' results also suggest that customers who purchase tickets closer to the departure date are less price sensitive and willing to pay higher seat selection fees. Interestingly, the authors also infer from their results that JetBlue's seat fees are underpriced in many markets and that an optimal static booking fee would increase revenues by 8%, while optimal dynamic fees—those that evolve over the booking period—would increase revenues by 10.2%. Finally, they suggest a corporate policy that would potentially increase the company's revenues by 12.8% through the combined use of a policy of maintaining seat fees at current levels, together with the reservation of certain rows of seats for premium customers, as several airlines, including Brazilian ones, have subsequently done.

Rouncivell, Timmis & Ison (2018) study the willingness to pay for seat selection on domestic flights in the United Kingdom, using a stated preference method. The authors investigate the relationship between passenger characteristics and opinions, with values they declared for paying seat reservation fees. The authors point out that consumer sensitivity to ticket prices, both for business and non-business travel (leisure and other personal reasons), is negatively correlated with the stated willingness to pay for seat reservations. On the other hand, customer perceptions of the airline's reputation and the convenience of its flight schedules are positively correlated with the willingness of non-business travelers to pay for seat selection. In addition, previous purchase of a seat selection product is strongly correlated with future willingness to pay for seat selection for both passenger segments. The authors interpret this result as stemming from the fact that, in this situation, consumers are better able to appreciate the benefits of their chosen seat based on their accumulated experience.

Shao, Kauermann & Smith (2020) study the decision to purchase seat selection and, given that decision, when and which seats are selected by passengers. To answer these questions, they use a dataset of 485,279 reservations on five intercontinental routes, extracted from the reservation history of a major European airline.

The authors find what they consider to be strong evidence of passenger purchasing behaviors of "avoiding middle seats" and "preferring front seats." Their results also suggest that the probability of purchasing seat selection depends on its price relative to the ticket price, the ticket distribution channel used, the advance purchase date, and seasonal effects.

Zhou et al. (2020) study the willingness to pay for seat selection in economy class on Chinese airlines. They use a mixed-methods approach, combining individual interviews with an online survey, to explore factors that influence the willingness of air consumers to pay for seat selection in economy class. The authors find that both intrinsic factors, such as trip duration, seat comfort, and convenience, and extrinsic factors, such as payment and consumption situations, have a significant impact on Chinese consumers' willingness to pay for seat selection.

Finally, one of the most radical proposals regarding changes to aircraft cabin layout is the idea of transporting passengers standing up. A concrete example of this approach is the Skyrider, whose concept will be described in the next section. One of the few studies analyzing this possibility is Romli et al. (2014), which investigates the feasibility of reducing operating costs and, consequently, ticket prices by increasing passenger capacity per flight. The study assumes that this expansion could dilute fixed costs and make air travel more affordable, especially in price-sensitive markets such as short-haul and low-cost travel. Based on a case study in the Malaysian market, the authors conclude that, although theoretically feasible on short flights, its adoption would depend on further analysis of safety, comfort, and structural impacts on aircraft.

## II.3. WILLINGNESS TO PAY FOR COMFORT AND AIRLINE PRODUCT DIFFERENTIATION

The study by Lee & Luengo-Prado (2004) is one of the central works on the economic value of comfort on board, particularly regarding seat pitch. The authors investigate whether passengers paid relatively higher fares after two initiatives implemented in 2000 by major US airlines: American Airlines' "More Room Throughout Coach" program and United Airlines' "Economy Plus" program. Both increased the space between rows in economy class, but with different strategies. American increased the pitch on all seats, while United applied the increase only to a few rows, targeting mainly corporate passengers and high-loyalty customers.

The analysis uses OD1A panel data covering 1998 to 2002 and approximately 1,000 pairs of airports with overlapping routes among the "Big Six." The dependent variable is the average round-trip fare. Control variables include market share, itinerary distance, origin airport share, flight frequency, corporate passenger proxy, direct flight supply, punctuality, financial leverage, and time trend. These controls reduce the influence of structural factors that differentiate companies, although the pitch effect is captured indirectly through pre- and post-dummies for American and United, given that the authors' database does not contain explicit measures of internal configuration.

The authors' fixed effects models per airport pair suggest results in line with the product differentiation literature. In the case of American Airlines, the overall increase in space did not translate into fare premiums. The estimated coefficients indicate a decline in American's relative premium after the intervention and an additional reduction in its fares compared to the pre-intervention period. For United Airlines, however, there was an increase in the relative premium after the adoption of Economy Plus, in the order of US$ 11 in the aggregate model, suggesting that the segmented strategy targeting less price-sensitive passengers was more effective in generating perceived differentiation.

The results show that willingness to pay for comfort is heterogeneous. Leisure passengers tend to prioritize price, while corporate passengers tend to value space and convenience more. Thus, United's superior performance is associated with a focus on higher-yield segments, consistent with models of spatial competition and segmentation by elasticity. The article points out that some of the differences found may result from other unmeasured service attributes and that the total impact on revenue also depends on possible occupancy gains. This suggests the need for research that incorporates operational and behavioral variables and more direct measures of internal configuration.

In summary, Lee & Luengo-Prado (2004) provide evidence that increasing seat pitch can generate fare differentiation when implemented in a segmented manner and targeted at high-value passengers. The study reinforces the relevance of preference heterogeneity and provides an important empirical and methodological basis for current analyses of cabin layout, comfort, and pricing.

In this study, as we will see, the research objective and quantitative approach are equivalent to those used by Lee & Luengo-Prado (2004), in the sense that we also seek to identify differences in airline ticket pricing associated with the internal configuration of the cabin. The main distinction lies in the fact that we have microdata with individual seat locations, which broadens the analytical scope and allows us to examine distributional effects within the aircraft itself. On the other hand, we do not have a before-and-after time experiment, as in the case of American and United's interventions. Our identification stems from a comparison between aircraft that adopted and did not adopt certain configurations, as will be detailed below. The objective remains essentially the same: to assess whether changes in the onboard layout generate systematic differences in the prices paid by passengers.

# I II. CASE STUDY

This section presents the case study used in the study, based on Brazilian domestic air transport. First, we describe the evolution of the cabin layout adopted by companies that operate and operated in the country, highlighting structural changes relevant to seat density and passenger space. Next, recent concepts and proposals for cabin configuration developed by the industry are introduced, which help to contextualize alternatives for space use and potential implications for comfort and costs. Finally, an exploratory inspection of alternative arrangements applied to typical domestic aircraft is carried out, with the aim of illustrating how different layout choices can affect the internal organization of the cabin. This approach provides the necessary background for the conceptual model and econometric analysis presented in the following sections.

## III.1. EVOLUTION OF CABIN LAYOUT CONFIGURATION IN BRAZIL

Air transport in Brazil has undergone numerous transformations over the last few decades. With the economic deregulation of the 1990s and early 2000s, airlines gained strategic freedom in pricing, flight frequencies, and route networks. Operators were allowed free mobility, so that they could start or end flights with great agility. The entry of new companies was facilitated, restricted only to technical certification criteria. The two biggest events in the sector since then were the entry of Gol and Azul in 2001 and 2008, respectively. In both cases, competition intensified in the first few years after entry, because of the newcomers' market penetration pricing, causing great optimism about the sector's growth prospects.

More important than the short-term effect of the new competition were the market positioning strategies observed. Gol began its operations strongly based on the low-cost business model inspired by the American Southwest Airlines, and, even with ups and downs in its market repositioning, managed to maintain this guideline over time. The presence of a low-cost rival led the incumbent TAM—now Latam—to compete on costs, where the exploitation of traffic density economies through larger aircraft and the densification of cabins with rows of seats was the norm. Similarly, with the entry of Azul, there was strong price competition, even though the newcomer initially operated from the secondary airport of Campinas/Viracopos, with Embraer narrowbody jets with lower seating capacity.

A few years into the 2010s, a new paradigm for competition among existing carriers emerged in the sector: the race to generate ancillary revenue by charging extra fees and customized fares in "packages" of attributes. This movement, which peaked with the introduction of baggage fees in 2017, began earlier, with the break from strategies purely focused on aircraft cabin densification, starting with the introduction of extra space in the first rows for passengers willing to pay for this amenity on board.

In November 2013, Gol introduced its "GOL+Conforto" seat on flights between Rio de Janeiro and São Paulo (Congonhas-Santos Dumont airports). The innovation consisted of offering, for an extra fee, on its Boeing 737-800NG aircraft, seat selection in seats with a larger pitch, 34 inches, compared to the 30 inches previously in effect, with 50% greater recline and a blocked middle seat. These features were available up to the seventh row of the aircraft, which now had a capacity of 177 seats[2] . From the end of 2013, the company expanded its operation of aircraft configured with GOL+Comfort seats, reaching 100% of its domestic network in October 2014[3] .

---

[2] Source: www.melhoresdestinos.com.br.
[3] Source: www.passageirodeprimeira.com,

| row | | W | | | G | | | |
|---|---|---|---|---|---|---|---|---|
| | ‡ | A | B | C | D | E | F | ‡ |
| 1 | | 1 | 1 | 1 | 1 | 1 | 1 | |
| 2 | | 1 | 1 | 1 | 1 | 1 | 1 | |
| 3 | | 1 | 1 | 1 | 1 | 1 | 1 | |
| 4 | | 1 | 1 | 1 | 1 | 1 | 1 | |
| 5 | | 1 | 1 | 1 | 1 | 1 | 1 | |
| 6 | | 1 | 1 | 1 | 1 | 1 | 1 | |
| 7 | | 1 | 1 | 1 | 1 | 1 | 1 | |
| 8 | | 1 | 1 | 1 | 1 | 1 | 1 | |
| 9 | | 1 | 1 | 1 | 1 | 1 | 1 | |
| 10 | / | 1 | 1 | 1 | 1 | 1 | 1 | \ |
| 11 | / | 1 | 1 | 1 | 1 | 1 | 1 | \ |
| 12 | / | 1 | 1 | 1 | 1 | 1 | 1 | \ |
| 13 | / | 0 | 0 | 0 | 0 | 0 | 0 | \ |
| 14 | / | 1 | 1 | 1 | 1 | 1 | 1 | \ |
| 15 | / | 1 | 1 | 1 | 1 | 1 | 1 | \ |
| 16 | / ‡ | 0 | 0 | 0 | 0 | 0 | 0 | ‡ \ |
| 17 | / ‡ | 1 | 1 | 1 | 1 | 1 | 1 | ‡ \ |
| 18 | / | 1 | 1 | 1 | 1 | 1 | 1 | \ |
| 19 | / | 1 | 1 | 1 | 1 | 1 | 1 | \ |
| 20 | / | 1 | 1 | 1 | 1 | 1 | 1 | \ |
| 21 | | 1 | 1 | 1 | 1 | 1 | 1 | |
| 22 | | 1 | 1 | 1 | 1 | 1 | 1 | |
| 23 | | 1 | 1 | 1 | 1 | 1 | 1 | |
| 24 | | 1 | 1 | 1 | 1 | 1 | 1 | |
| 25 | | 1 | 1 | 1 | 1 | 1 | 1 | |
| 26 | | 1 | 1 | 1 | 1 | 1 | 1 | |
| 27 | | 1 | 1 | 1 | 1 | 1 | 1 | |
| 28 | | 1 | 1 | 1 | 1 | 1 | 1 | |
| 29 | | 1 | 1 | 1 | 1 | 1 | 1 | |
| 30 | | 1 | 1 | 1 | 1 | 1 | 1 | |
| 31 | | 1 | 1 | 1 | 1 | 1 | 1 | |
| 32 | | 1 | 1 | 1 | | | | |

Boeing 737-800 (Gol) — 177 SEATS

| row | | W | | | G | | | |
|---|---|---|---|---|---|---|---|---|
| | ‡ | A | B | C | D | E | F | ‡ |
| 1 | | 1 | 1 | 1 | 1 | 1 | 1 | |
| 2 | | 1 | 1 | 1 | 1 | 1 | 1 | |
| 3 | | 1 | 1 | 1 | 1 | 1 | 1 | |
| 4 | | 1 | 1 | 1 | 1 | 1 | 1 | |
| 5 | | 1 | 1 | 1 | 1 | 1 | 1 | |
| 6 | | 1 | 1 | 1 | 1 | 1 | 1 | |
| 7 | | 1 | 1 | 1 | 1 | 1 | 1 | |
| 8 | | 1 | 1 | 1 | 1 | 1 | 1 | |
| 9 | | 1 | 1 | 1 | 1 | 1 | 1 | |
| 10 | / | 1 | 1 | 1 | 1 | 1 | 1 | \ |
| 11 | / | 1 | 1 | 1 | 1 | 1 | 1 | \ |
| 12 | / | 1 | 1 | 1 | 1 | 1 | 1 | \ |
| 13 | / | 0 | 0 | 0 | 0 | 0 | 0 | \ |
| 14 | / | 1 | 1 | 1 | 1 | 1 | 1 | \ |
| 15 | / | 1 | 1 | 1 | 1 | 1 | 1 | \ |
| 16 | / ‡ | 0 | 1 | 1 | 1 | 1 | 0 | ‡ \ |
| 17 | / | 1 | 1 | 1 | 1 | 1 | 1 | \ |
| 18 | / | 1 | 1 | 1 | 1 | 1 | 1 | \ |
| 19 | / | 1 | 1 | 1 | 1 | 1 | 1 | \ |
| 20 | / | 1 | 1 | 1 | 1 | 1 | 1 | \ |
| 21 | | 1 | 1 | 1 | 1 | 1 | 1 | |
| 22 | | 1 | 1 | 1 | 1 | 1 | 1 | |
| 23 | | 1 | 1 | 1 | 1 | 1 | 1 | |
| 24 | | 1 | 1 | 1 | 1 | 1 | 1 | |
| 25 | | 1 | 1 | 1 | 1 | 1 | 1 | |
| 26 | | 1 | 1 | 1 | 1 | 1 | 1 | |
| 27 | | 1 | 1 | 1 | 1 | 1 | 1 | |
| 28 | | 1 | 1 | 1 | 1 | 1 | 1 | |
| 29 | | 1 | 1 | 1 | 1 | 1 | 1 | |
| 30 | | 1 | 1 | 1 | 1 | 1 | 1 | |
| 31 | | 1 | 1 | 1 | 1 | 1 | 1 | |

Boeing 737-800 (Gol) — 178 SEATS

*Symbols: "G" (galley); "W" (toilet); "/" and "\" (wing); "‡" emergency exit; "A", "B", "C" and "D" (seat position); "row" (row); "0" (seat does not exist); "1" seat exists"; "1" seat exists with greater pitch". Notes: maps are not to scale; sizes of seat rectangles do not reflect seat pitch and width proportions; approximate positions of aircraft cabin components; a row containing only "0" indicates that there is no space in that row. Sources: PANROTAS Guide 2014, nos. 490-493 and website www.web.archive.org/web/www.seatguru.com (year 2014).*

**Figure 1 - Seat map - Gol aircraft (2014) - examples**

| ATR-72-600 (Azul) |||||||
|---|---|---|---|---|---|---|
| 68 SEATS |||||||
| row | | G | | G | | |
| | ‡ | A | B | C | D | ‡ |
| 1 | | 1 | 1 | 1 | 1 | |
| 2 | | 1 | 1 | 1 | 1 | |
| 3 | | 1 | 1 | 1 | 1 | |
| 4 | | 1 | 1 | 1 | 1 | |
| 5 | | 1 | 1 | 1 | 1 | |
| 6 | / | 1 | 1 | 1 | 1 | \ |
| 7 | / | 1 | 1 | 1 | 1 | \ |
| 8 | / | 1 | 1 | 1 | 1 | \ |
| 9 | / | 1 | 1 | 1 | 1 | \ |
| 10 | / | 1 | 1 | 1 | 1 | \ |
| 11 | / | 1 | 1 | 1 | 1 | \ |
| 12 | / | 1 | 1 | 1 | 1 | \ |
| 13 | | 1 | 1 | 1 | 1 | |
| 14 | | 1 | 1 | 1 | 1 | |
| 15 | | 1 | 1 | 1 | 1 | |
| 16 | | 1 | 1 | 1 | 1 | |
| 17 | | 1 | 1 | 1 | 1 | |
| | ‡ | | | | | ‡ |
| | | W | | G | | |

| Embraer E-195 (Azul) |||||||
|---|---|---|---|---|---|---|
| 118 SEATS |||||||
| row | | W | | G | | |
| | ‡ | A | B | C | D | ‡ |
| 1 | | 0 | 0 | 1 | 1 | |
| 2 | | 1 | 1 | 1 | 1 | |
| 3 | | 1 | 1 | 1 | 1 | |
| 4 | | 1 | 1 | 1 | 1 | |
| 5 | | 1 | 1 | 1 | 1 | |
| 6 | | 1 | 1 | 1 | 1 | |
| 7 | | 1 | 1 | 1 | 1 | |
| 8 | | 1 | 1 | 1 | 1 | |
| 9 | | 1 | 1 | 1 | 1 | |
| 10 | / | 1 | 1 | 1 | 1 | \ |
| 11 | / | 1 | 1 | 1 | 1 | \ |
| 12 | / | 1 | 1 | 1 | 1 | \ |
| 13 | / | 1 | 1 | 1 | 1 | \ |
| 14 | / ‡ | 1 | 1 | 1 | 1 | ‡ \ |
| 15 | / | 1 | 1 | 1 | 1 | \ |
| 16 | / | 1 | 1 | 1 | 1 | \ |
| 17 | / | 1 | 1 | 1 | 1 | \ |
| 18 | / | 1 | 1 | 1 | 1 | \ |
| 19 | / | 1 | 1 | 1 | 1 | \ |
| 20 | / | 1 | 1 | 1 | 1 | \ |
| 21 | | 1 | 1 | 1 | 1 | |
| 22 | | 1 | 1 | 1 | 1 | |
| 23 | | 1 | 1 | 1 | 1 | |
| 24 | | 1 | 1 | 1 | 1 | |
| 25 | | 1 | 1 | 1 | 1 | |
| 26 | | 1 | 1 | 1 | 1 | |
| 27 | | 1 | 1 | 1 | 1 | |
| 28 | | 1 | 1 | 1 | 1 | |
| 29 | | 1 | 1 | 1 | 1 | |
| 30 | | 1 | 1 | 1 | 1 | |
| | ‡ | | | | | ‡ |
| | | W | | G | | |

*Symbols: "G" (galley); "W" (toilet); "/" and "\" (wing); "‡" emergency exit; "A", "B", "C" and "D" (seat position); "row" (row); "0" (seat does not exist); "1" seat exists"; "1" seat exists with greater pitch". Notes: maps are not to scale; rectangle sizes indicating seats do not reflect their pitch and width proportions; approximate positions of aircraft cabin components; a row containing only "0" indicates that there is no space in that row. Sources: PANROTAS Guide 2014, nos. 490-493 and website www.web.archive.org/web/www.seatguru.com (year 2014).*

**Figure 2 - Seat map - Azul aircraft (2014) - examples**

| Airbus A319 (Avianca Brasil) 132 SEATS | | | | | | | | | | Airbus A320 (Avianca Brasil) 162 SEATS | | | | | | | | | |
|---|---|---|---|---|---|---|---|---|---|---|---|---|---|---|---|---|---|---|---|
| row | | | W | | | G | | | | row | | | W | | | G | | | |
| | ‡ | A | B | C | D | E | F | ‡ | | | ‡ | A | B | C | D | E | F | ‡ | |
| 1 | | 1 | 1 | 1 | 1 | 1 | 1 | | | 1 | | 1 | 1 | 1 | 1 | 1 | 1 | | |
| 2 | | 1 | 1 | 1 | 1 | 1 | 1 | | | 2 | | 1 | 1 | 1 | 1 | 1 | 1 | | |
| 3 | | 1 | 1 | 1 | 1 | 1 | 1 | | | 3 | | 1 | 1 | 1 | 1 | 1 | 1 | | |
| 4 | | 1 | 1 | 1 | 1 | 1 | 1 | | | 4 | | 1 | 1 | 1 | 1 | 1 | 1 | | |
| 5 | | 1 | 1 | 1 | 1 | 1 | 1 | | | 5 | | 1 | 1 | 1 | 1 | 1 | 1 | | |
| 6 | | 1 | 1 | 1 | 1 | 1 | 1 | | | 6 | | 1 | 1 | 1 | 1 | 1 | 1 | | |
| 7 | / | 1 | 1 | 1 | 1 | 1 | 1 | | \ | 7 | | 1 | 1 | 1 | 1 | 1 | 1 | | |
| 8 | / | 1 | 1 | 1 | 1 | 1 | 1 | | \ | 8 | | 1 | 1 | 1 | 1 | 1 | 1 | | |
| 9 | / | 1 | 1 | 1 | 1 | 1 | 1 | | \ | 9 | / | 1 | 1 | 1 | 1 | 1 | 1 | | \ |
| 10 | / ‡ | 1 | 1 | 1 | 1 | 1 | 1 | ‡ | \ | 10 | / | 1 | 1 | 1 | 1 | 1 | 1 | | \ |
| 11 | / | 1 | 1 | 1 | 1 | 1 | 1 | | \ | 11 | / ‡ | 1 | 1 | 1 | 1 | 1 | 1 | ‡ | \ |
| 12 | / | 1 | 1 | 1 | 1 | 1 | 1 | | \ | 12 | / ‡ | 1 | 1 | 1 | 1 | 1 | 1 | ‡ | \ |
| 13 | / | 0 | 0 | 0 | 0 | 0 | 0 | | \ | 13 | / | 0 | 0 | 0 | 0 | 0 | 0 | | \ |
| 14 | / | 1 | 1 | 1 | 1 | 1 | 1 | | \ | 14 | / | 1 | 1 | 1 | 1 | 1 | 1 | | \ |
| 15 | | 1 | 1 | 1 | 1 | 1 | 1 | | | 15 | / | 1 | 1 | 1 | 1 | 1 | 1 | | \ |
| 16 | | 1 | 1 | 1 | 1 | 1 | 1 | | | 16 | / | 1 | 1 | 1 | 1 | 1 | 1 | | \ |
| 17 | | 1 | 1 | 1 | 1 | 1 | 1 | | | 17 | / | 1 | 1 | 1 | 1 | 1 | 1 | | \ |
| 18 | | 1 | 1 | 1 | 1 | 1 | 1 | | | 18 | / | 1 | 1 | 1 | 1 | 1 | 1 | | \ |
| 19 | | 1 | 1 | 1 | 1 | 1 | 1 | | | 19 | | 1 | 1 | 1 | 1 | 1 | 1 | | |
| 20 | | 1 | 1 | 1 | 1 | 1 | 1 | | | 20 | | 1 | 1 | 1 | 1 | 1 | 1 | | |
| 21 | | 1 | 1 | 1 | 1 | 1 | 1 | | | 21 | | 1 | 1 | 1 | 1 | 1 | 1 | | |
| 22 | | 1 | 1 | 1 | 1 | 1 | 1 | | | 22 | | 1 | 1 | 1 | 1 | 1 | 1 | | |
| 23 | | 1 | 1 | 1 | 1 | 1 | 1 | | | 23 | | 1 | 1 | 1 | 1 | 1 | 1 | | |
| | | | W | | | W | | | | 24 | | 1 | 1 | 1 | 1 | 1 | 1 | | |
| | ‡ | | | | | | | ‡ | | 25 | | 1 | 1 | 1 | 1 | 1 | 1 | | |
| | | | | G | | | | | | 26 | | 1 | 1 | 1 | 1 | 1 | 1 | | |
| | | | | | | | | | | 27 | | 1 | 1 | 1 | 1 | 1 | 1 | | |
| | | | | | | | | | | 28 | | 1 | 1 | 1 | 1 | 1 | 1 | | |
| | | | | | | | | | | | | | W | | | W | | | |
| | | | | | | | | | | | ‡ | | | | | | | ‡ | |
| | | | | | | | | | | | | | | G | | | | | |

*Symbols: "G" (galley); "W" (toilet); "/" and "\" (wing); "‡" emergency exit; "A," "B," "C," and "D" (seat position); "row"; "0" (seat does not exist); "1" seat exists. Notes: maps are not to scale; sizes of seat rectangles do not reflect seat pitch and width proportions; approximate positions of aircraft cabin components; a row containing only "0" indicates that there is no space in that row. Sources: PANROTAS Guide 2014, nos. 490-493 and website www.web.archive.org/web/www.seatguru.com (year 2014).*

**Figure 3 - Seat map - Avianca aircraft (2014) - examples**

| row | | W | | | G | | | |
|---|---|---|---|---|---|---|---|---|
| | ‡ | A | B | C | D | E | F | ‡ |
| | | Airbus A320 (TAM) 156 SEATS | | | | | | |
| 1 | | <u>1</u> | <u>0</u> | <u>1</u> | <u>1</u> | <u>0</u> | <u>1</u> | |
| 2 | | <u>1</u> | <u>0</u> | <u>1</u> | <u>1</u> | <u>0</u> | <u>1</u> | |
| 3 | | <u>1</u> | <u>0</u> | <u>1</u> | <u>1</u> | <u>0</u> | <u>1</u> | |
| 4 | | 0 | 0 | 0 | 0 | 0 | 0 | |
| 5 | | 1 | 1 | 1 | 1 | 1 | 1 | |
| 6 | | 1 | 1 | 1 | 1 | 1 | 1 | |
| 7 | | 1 | 1 | 1 | 1 | 1 | 1 | |
| 8 | | 1 | 1 | 1 | 1 | 1 | 1 | |
| 9 | / | 1 | 1 | 1 | 1 | 1 | 1 | \ |
| 10 | / | 1 | 1 | 1 | 1 | 1 | 1 | \ |
| 11 | / | 1 | 1 | 1 | 1 | 1 | 1 | \ |
| 12 | / ‡ | 1 | 1 | 1 | 1 | 1 | 1 | ‡ \ |
| 13 | / | 1 | 1 | 1 | 1 | 1 | 1 | \ |
| 14 | / | 1 | 1 | 1 | 1 | 1 | 1 | \ |
| 15 | / | 1 | 1 | 1 | 1 | 1 | 1 | \ |
| 16 | / | 1 | 1 | 1 | 1 | 1 | 1 | \ |
| 17 | / | 1 | 1 | 1 | 1 | 1 | 1 | \ |
| 18 | / | 1 | 1 | 1 | 1 | 1 | 1 | \ |
| 19 | | 1 | 1 | 1 | 1 | 1 | 1 | |
| 20 | | 1 | 1 | 1 | 1 | 1 | 1 | |
| 21 | | 1 | 1 | 1 | 1 | 1 | 1 | |
| 22 | | 1 | 1 | 1 | 1 | 1 | 1 | |
| 23 | | 1 | 1 | 1 | 1 | 1 | 1 | |
| 24 | | 1 | 1 | 1 | 1 | 1 | 1 | |
| 25 | | 1 | 1 | 1 | 1 | 1 | 1 | |
| 26 | | 1 | 1 | 1 | 1 | 1 | 1 | |
| 27 | | 1 | 1 | 1 | 1 | 1 | 1 | |
| 28 | | 1 | 1 | 1 | 1 | 1 | 1 | |
| | | W | | | W | | | |
| | ‡ | | | | | | | ‡ |
| | | | | G | | | | |

| row | | W | | | G | | | |
|---|---|---|---|---|---|---|---|---|
| | ‡ | A | B | C | D | E | F | ‡ |
| | | Airbus A320 (TAM) 174 SEATS | | | | | | |
| 1 | | 1 | 1 | 1 | 1 | 1 | 1 | |
| 2 | | 1 | 1 | 1 | 1 | 1 | 1 | |
| 3 | | 1 | 1 | 1 | 1 | 1 | 1 | |
| 4 | | 1 | 1 | 1 | 1 | 1 | 1 | |
| 5 | | 1 | 1 | 1 | 1 | 1 | 1 | |
| 6 | | 1 | 1 | 1 | 1 | 1 | 1 | |
| 7 | | 1 | 1 | 1 | 1 | 1 | 1 | |
| 8 | | 1 | 1 | 1 | 1 | 1 | 1 | |
| 9 | | 1 | 1 | 1 | 1 | 1 | 1 | |
| 10 | / | 1 | 1 | 1 | 1 | 1 | 1 | \ |
| 11 | / | 1 | 1 | 1 | 1 | 1 | 1 | \ |
| 12 | / ‡ | 1 | 1 | 1 | 1 | 1 | 1 | ‡ \ |
| 13 | / | 1 | 1 | 1 | 1 | 1 | 1 | \ |
| 14 | / | 1 | 1 | 1 | 1 | 1 | 1 | \ |
| 15 | / | 1 | 1 | 1 | 1 | 1 | 1 | \ |
| 16 | / | 1 | 1 | 1 | 1 | 1 | 1 | \ |
| 17 | / | 1 | 1 | 1 | 1 | 1 | 1 | \ |
| 18 | / | 1 | 1 | 1 | 1 | 1 | 1 | \ |
| 19 | / | 1 | 1 | 1 | 1 | 1 | 1 | \ |
| 20 | / | 1 | 1 | 1 | 1 | 1 | 1 | \ |
| 21 | | 1 | 1 | 1 | 1 | 1 | 1 | |
| 22 | | 1 | 1 | 1 | 1 | 1 | 1 | |
| 23 | | 1 | 1 | 1 | 1 | 1 | 1 | |
| 24 | | 1 | 1 | 1 | 1 | 1 | 1 | |
| 25 | | 1 | 1 | 1 | 1 | 1 | 1 | |
| 26 | | 1 | 1 | 1 | 1 | 1 | 1 | |
| 27 | | 1 | 1 | 1 | 1 | 1 | 1 | |
| 28 | | 1 | 1 | 1 | 1 | 1 | 1 | |
| 29 | | 1 | 1 | 1 | 1 | 1 | 1 | |
| | | W | | | W | | | |
| | ‡ | | | | | | | ‡ |
| | | | | G | | | | |

*Symbols: "G" (galley); "W" (toilet); "/" and "\" (wing); "‡" emergency exit; "A", "B", "C" and "D" (seat position); "row" (row); "0" (seat does not exist); "1" seat exists"; "<u>1</u>" seat exists with greater pitch". Notes: maps are not to scale; sizes of rectangles indicating seats do not reflect their pitch and width proportions; approximate positions of aircraft cabin components; a row containing only "0" indicates that there is no space in that row. Sources: PANROTAS Guide 2014, nos. 490-493 and website www.web.archive.org/web/www.seatguru.com (year 2014).*

**Figure 4 - Seat map - TAM aircraft (2014) - examples**

In what follows, we present a sequence of figures with illustrative maps of the layout configurations of the aircraft in operation during the period analyzed. The diagrams are not to scale and use symbols adopted by the authors, detailed in the footer of each image, indicating essential components of the cabin, such as galley, bathrooms, wings, emergency exits, last row, and specific seat positions. Particular attention should be paid to the cells marked with the number 1, which may appear with or without underlining, with the underlining identifying the existence of a seat with extra pitch offered by the airline. These maps, constructed from analyses of maps from historical editions of the Panrotas Guide, are representative in nature and serve to visualize relevant structural differences between aircraft and airlines at the time the GOL+Conforto product was introduced.

1 shows two possible configurations of Gol's Boeing 737-800, an aircraft manufactured by Boeing and widely used in the Brazilian domestic market. On the left is the 177-seat layout, in which the first seven rows have seats with extra space, indicated by the underlined "1". On the right is the 178-seat version, without any extra space, reflecting the standard high-density configuration adopted in part of the fleet during the period analyzed.

Figures 2, 3, and 4 illustrate examples of aircraft seat maps from the main airlines at the time of the introduction of the GOL+Comfort seat. In Figure 2, on the left is Azul's ATR-72-600, manufactured by ATR, with 68 seats, and on the right is the Embraer E-195, with 118 seats. Figure 3 shows Avianca Brasil's Airbus A319 with 132 seats on the left and Avianca Brasil's Airbus A320 with 162 seats on the right. Finally, Figure 4 shows two layouts of the Airbus A320 operated by TAM, now Latam, with the aircraft on the left configured with 156 seats and the one on the right configured with 174 seats.

### III.2. Innovative cabin layout concepts

The low-cost business model, associated with airlines such as Southwest Airlines in the United States and Ryanair in the United Kingdom, is based on operating aircraft configured with high seat density. The aim is to accommodate as many passengers per flight as possible and thereby minimize unit costs per passenger. Thus, one of the biggest innovations introduced by this business model was the controversial reduction in seat pitch by adding rows to aircraft to achieve their maximum transport configuration, within the limits imposed by regulation. Signaling a deepening of these strategies, in 2010, Ryanair conducted a survey of 120,000 people, in a study that concluded that one-third of them would consider purchasing tickets for flights in "vertical seats" on aircraft if the tickets were free, while 42% said they would use them if the price of the tickets were half that of a ticket for flights in conventional seats[4]. As we will discuss later, the "vertical seat" is one of the innovative cabin design concepts under consideration in the industry, which allows for up to a 20% increase in seating capacity on a commercial jet[5].

On the other hand, seat manufacturer Zephyr Aerospace conducted a survey in the late 2010s and found that 70% of economy class travelers would be willing to trade some of the perks associated with extra fees then offered by airlines, such as in-flight meal packages and extra baggage, for the possibility of "lying down and sleeping"[6]. An important aspect of the innovative efforts of aircraft seat manufacturers and designers is to offer solutions that optimize the number of passengers carried on a flight to meet the demand of low-cost airlines and their rivals that need to be more competitive, but also to offer possibilities for greater comfort on board to generate ancillary revenue and overall profitability.

Below, we present some of the innovative passenger aircraft seat configuration concepts presented by developers in recent years. Some of these concepts have been under discussion since the late 2000s. It is known that many of them have already been patented[7], although they have not been certified by aviation authorities. The innovative concepts vary in terms of their likelihood of being accepted on a large scale by airlines and the public, given their relative advantages and disadvantages. Some are so futuristic that it is difficult to clearly anticipate the profile of passengers willing to accept them and how they would fit into the commercial strategies of airlines. It is possible that many will never actually obtain certification, given the strict flight safety requirements that must be met, for example, with issues related to procedures in the event of turbulence or evacuation of the aircraft in an emergency. As an illustration of this rigor in the evaluation of new aeronautical products, we can highlight a statement made by an Airbus representative in 2014 who suggested that "*many, if not most, patent projects [relating to aircraft seats] will never be developed*"[8] . However, it is necessary to investigate their effects to anticipate future paradigm shifts, which are generally disruptive and difficult to predict based on experience.

---

[4] www.dailymail.co.uk.
[5] www.aviointeriors.it/2018/press/aviointeriors-skyrider-2-0. See our discussion below on the "Skyrider."
[6] www.dailyhive.com/mapped/double-decker-zephyr-seat-airplane.
[7] www.today.com/money/stacked-squatting-squished-7-scary-airline-seat-patents-t52241.

It is necessary to highlight the uncertainty associated with the impact of these concepts on an airline's operating costs, including fuel consumption, maintenance, and other components. On the other hand, they all entail significant changes in seat density, with effects on unit costs, comfort, prices, segmentation, and airline profitability. With changes in the number of rows and seats, it becomes possible to discuss some of their possible operational and commercial impacts on air transport in a more informed manner. It is important to note, however, that certain solutions are more applicable to the international segment than to the domestic segment, given the differences in operations, market, and passenger profiles.

The first seat design concept is "Morph Seating" by Seymour Powell (Figure 5).

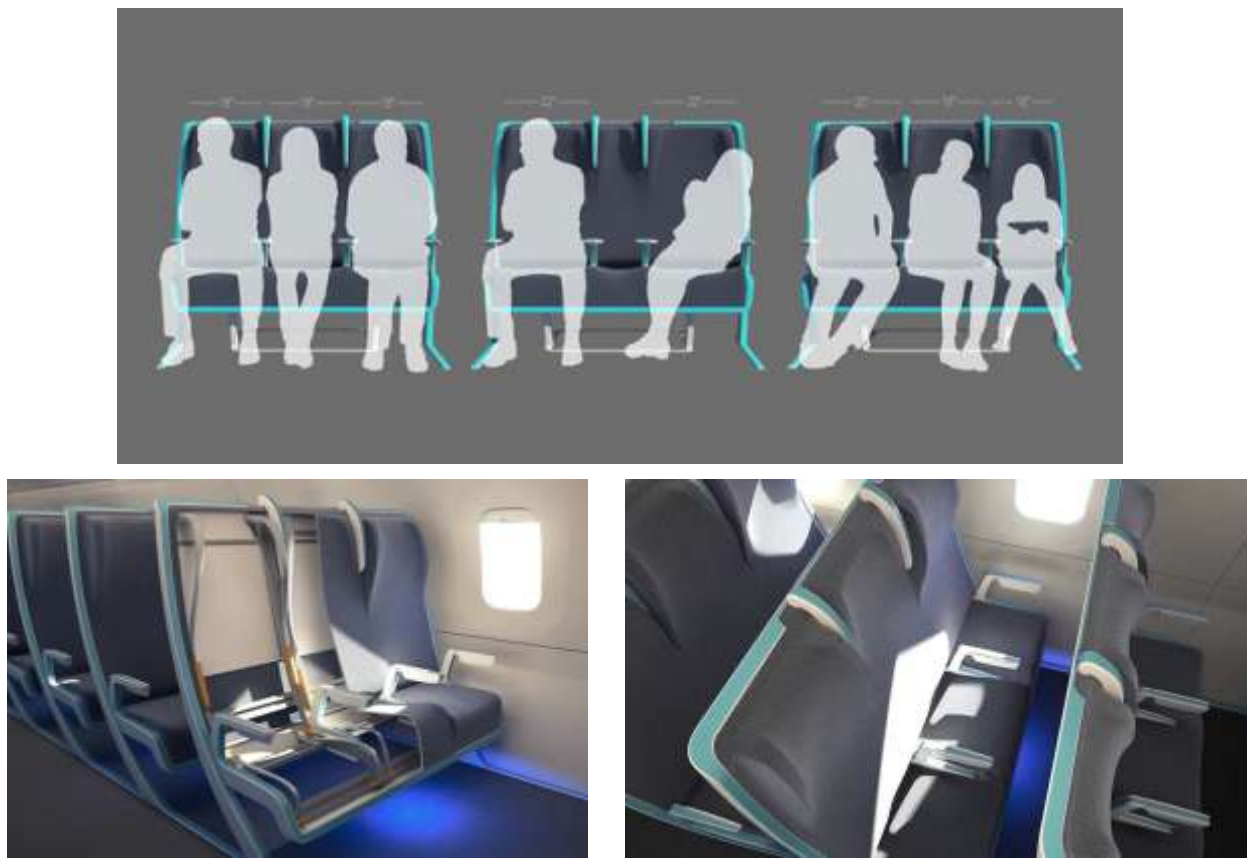

*Sources: www.seymourpowell.com; www.youtube.com/watch?v=X8REY-oXl3U.*

**Figure 5 - "Morph Seating" concept (Seymour Powell)**

The main idea behind "Morph Seating" is that, instead of having a trio of individual seats, there is a type of bench made from several pieces of fabric and foam. A single piece of fabric is stretched to form the seats and another forms the backrest. The individual seats are defined by armrests and dividers that secure the fabric. The concept is based on the use of stretched fabric and movable supports that allow passengers to customize their seats and enable airlines to charge for additional seat width among the amenities offered at the time of booking. The flexibility of width is its main distinguishing feature.

From a capacity and density standpoint, Morph Seating could allow for width adjustments based on the passenger composition of each flight, increasing density when there is a higher proportion of travelers willing to accept narrower seats and reducing it when the focus is on comfort, which makes it attractive to segments with a greater willingness to pay, such as frequent business travelers, corporate passengers, and premium customers on medium-haul routes. This flexibility would allow airlines to test different combinations of comfort and capacity, applying different fares based on width and better managing the balance between revenue per seat and passenger satisfaction. This same flexibility, however, can introduce commercial and operational complexity, as well as the risk of frustration if the perception of fairness in space distribution is affected, if the contracted width does not correspond to the actual experience, or if conflicts arise between neighboring passengers with differing preferences. The need for physical adjustments to supports and partitions and the requirement for standardization of safety, cleaning, and maintenance may reduce some of the expected gains, so that the concept tends to be more suited to niches or hybrid cabins with zones of greater comfort than to a solution that can be widely applied to economy class.

---

[8] www.newsweek.com/saddle-seat-economy-flights-could-have-you-standing-la-new-york-888943.

Another concept introduced recently is that of Thompson Aero Seating, called the "Cozy Suite" (Figure 6).

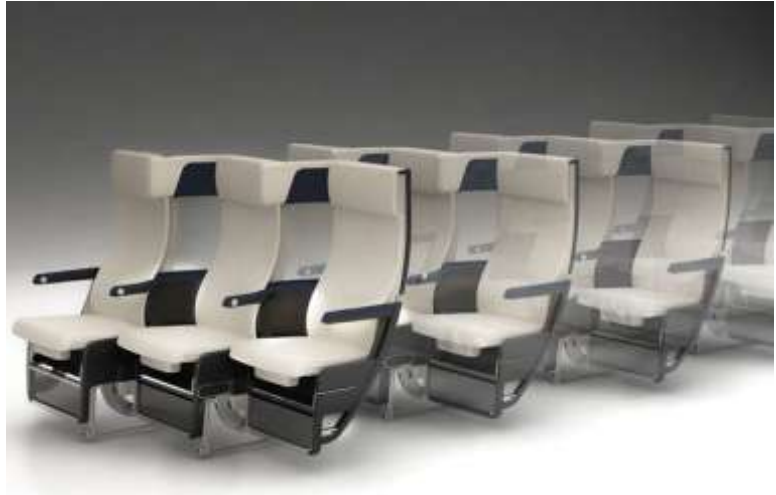

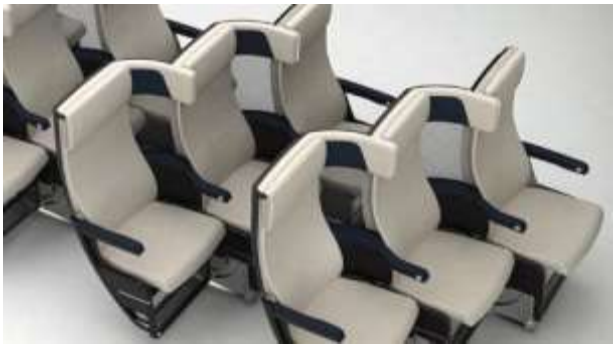 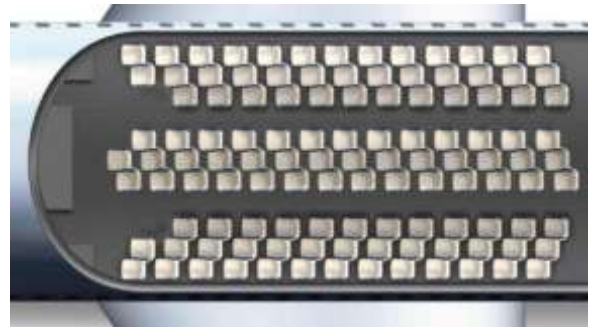

*Source: www.thompsonaero.com*
.

Figure 6 - "Cozy Suite" concept (Thompson Aero Seating)

This is a seat design that incorporates a second side headrest, in addition to individual armrests. In theory, it would offer a solution to the conflict generated by the shared use of armrests, as well as providing a wider seat and easier access to the aisles. The developer indicates the possibility of increasing seat density. In terms of comparative advantages, the concept clearly benefits the middle seat passenger compared to the layout currently used in commercial aviation.

The "Cozy Suite" has direct implications for capacity, density, and market segmentation. By including a second side headrest and reorganizing the relative space between the three passengers, the concept can increase density without compromising the occupant experience. This reorganization creates space for new product levels within economy class, allowing for the marketing of zones with greater privacy and ergonomics or the application of differentiated fares to the redesigned middle seat. In terms of segmentation, there are possibilities for targeting travelers who value lateral privacy and corporate passengers who prefer a seat that reduces lateral body movement during the flight. The promise of increased density, however, requires caution, as structural rearrangements can create challenges in terms of certification, ergonomics, and evacuation, as well as conflicts in the perception of fairness among passengers who pay for attributes that are difficult to standardize. There is also a risk that the gain in comfort will be perceived as excessively asymmetrical, disproportionately favoring one type of seat and creating comparisons that may generate dissatisfaction, which limits the widespread adoption of the concept and points to its use in specific segments or in more qualified areas within economy class.

Figure 7 presents the concept of the "Air Lair." The Air Lair is a two-story "individual cocoon" space for passengers. It is an ergonomically designed flat seat, like a low car seat, to provide comfort to passengers. The manufacturer estimates that the configuration accommodates 30% more passengers within the same cabin space[9] . Within the isolated cocoon environment, passengers can adjust control their own personal space without affecting other passengers. The manufacturer proposes the use of strategically placed lighting to create an environment that encourages the use of onboard entertainment equipment, thereby generating additional ancillary revenue.

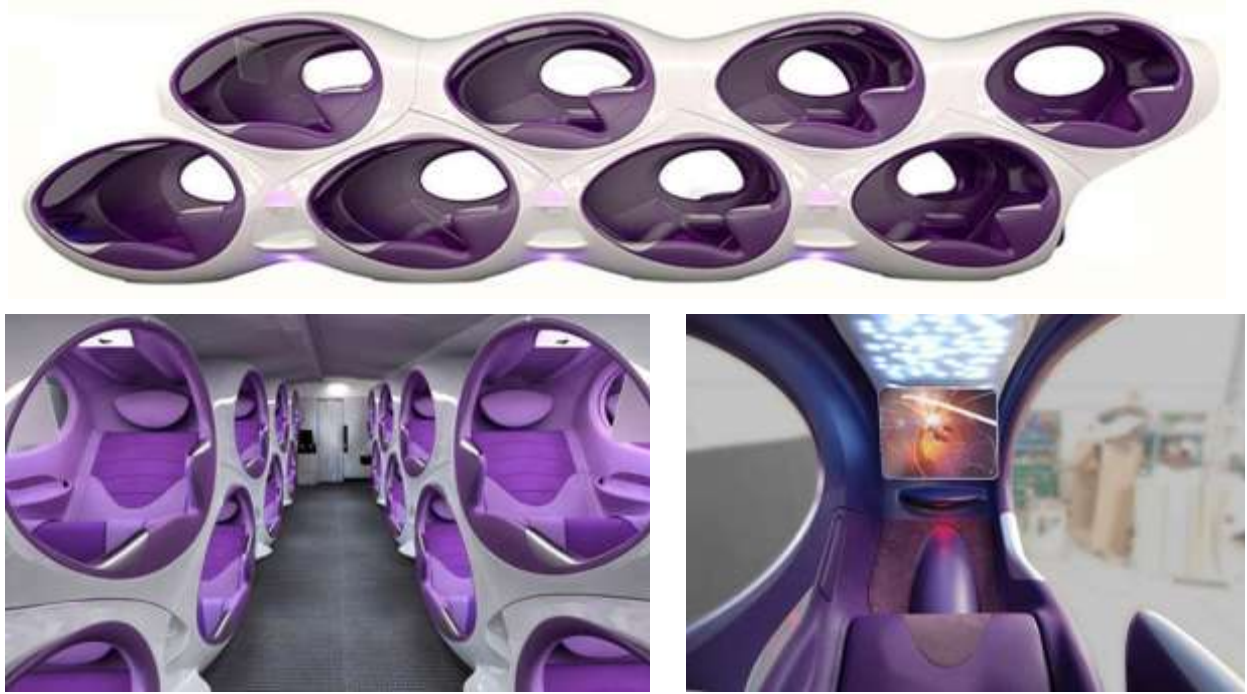

*Sources: www.collabcubed.com/2012/06/13/air-lair-pod-business-class-seats;*
*www.dailymail.co.uk/travel/travel_news/article-2839713/Take-skies-double-decker-sleep-pods-Futuristic-stacked-cocoons-FINALLY-air-passengers-privacy.html.*

**Figure 7 - "Air Lair" concept (Factorydesign)**

The "Air Lair" radically changes the relationship between capacity, density, and segmentation by transforming the space into stacked pods on two levels and offering individual isolation that reduces conflicts over space and interaction between passengers. The ability to accommodate around 30 percent more passengers creates a significant density advantage, especially on long-haul international routes, where the cost of space per passenger is high. For the airline, the model allows for deeper segmentation, with pods differentiated by level, position, privacy, lighting, and entertainment integration, bringing the cabin logic closer to the modular products found in premium services. The pod format, however, may create barriers to acceptance among passengers who reject more enclosed environments, as well as regulatory challenges related to evacuation, circulation, and accessibility. There is also a risk of perceived inequality if differences between pods are difficult to justify through fares or if increased density compromises the overall feeling of space. Thus, the concept tends to align with specific niches or mixed cabins aimed at maximizing revenue on international flights, rather than being configured as a broadly applicable solution for economy class.

---

[9] www.factorydesign.co.uk/aviation/case-study-aviation-air-lair,

Figure 8 shows images of the "StepSeat" from Jacob-Innovations. The StepSeat consists of alternating raised seats to take advantage of vertical space inside aircraft using steps. The idea behind the concept is to provide greater comfort to passengers while maintaining the same seat density on board[10].

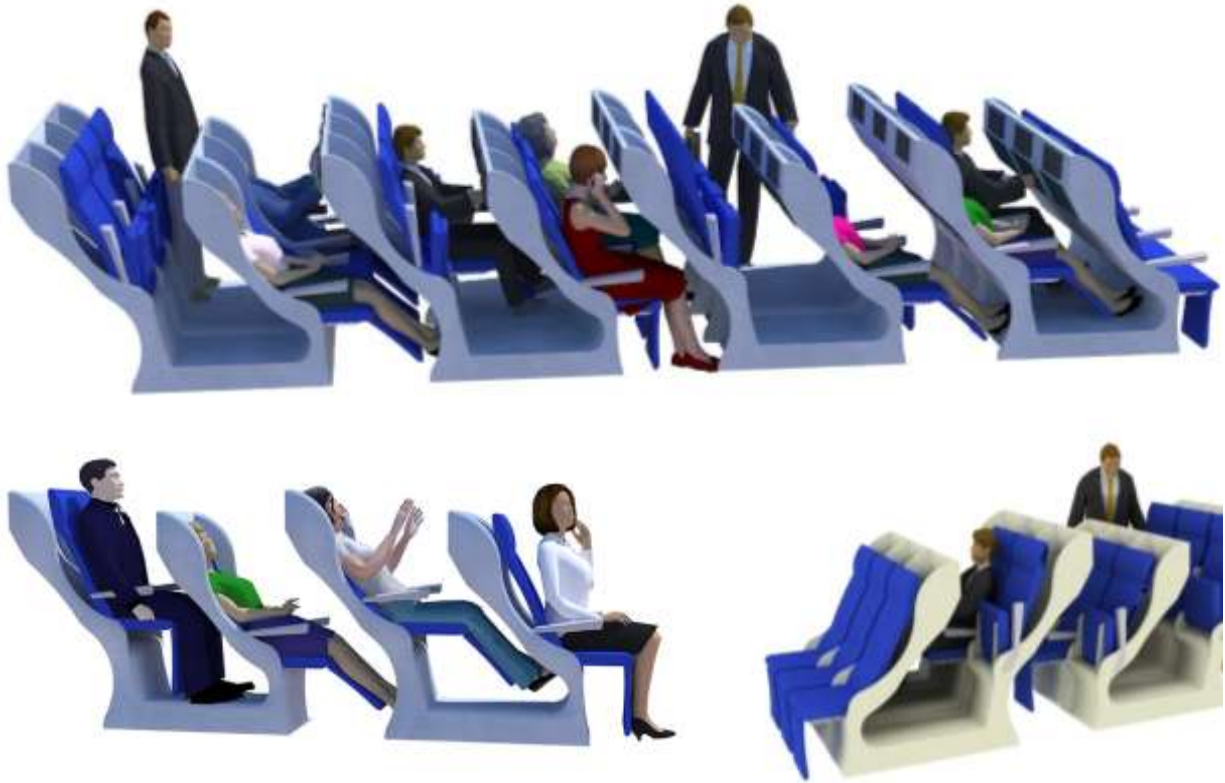

*Source: www.jacob-innovations.com.*

**Figure 8 - "StepSeat" concept (Jacob-Innovations)**

The "StepSeat" changes the spatial organization by introducing alternating raised seats, utilizing vertical volume without changing the total number of seats. In principle, this would increase relative comfort per passenger without sacrificing density, creating a product that could occupy an intermediate position between standard seating and more expensive comfort alternatives. For market segmentation, the airline could explore different categories based on height, privacy, and ergonomics, targeting the product at passengers who value a sense of personal space, business travelers on medium-haul routes, or customers willing to pay for a less conventional experience. A critical analysis, however, indicates challenges such as possible evacuation difficulties, safety in the use of steps, standardization of service, and public acceptance, especially among passengers with reduced mobility or aversion to elevated structures. There is also the risk of a very heterogeneous perception of comfort, which makes the product more suitable for niches or restricted sections of the cabin, especially in international operations, where there is greater tolerance for structural innovations and a greater willingness to pay for differentiation.

Another concept from Jacob-Innovations that uses steps to capture unused vertical space in current seats is FlexSeat. (Figure 9). FlexSeat uses the idea of a "double-decker" airplane, a concept already used in the largest Boeing 747 and Airbus A380 airplanes, as well as some bus and train models. The manufacturer emphasizes not only the feature of seats with greater recline than current ones, but also features superior even to conventional Business Class, such as privacy, the possibility of traveling with babies, and space for larger carry-on luggage. Additionally, the concept allows for an "instant" conversion from Business Class to Economy Class[11] by adjusting the steps to the upper level. This feature would allow airlines to sell tickets for both classes according to demand.

---

[10] www.jacob-innovations.com/Economy-Comfort.html.
[11] www.jacob-innovations.com/FlexSeat.html.

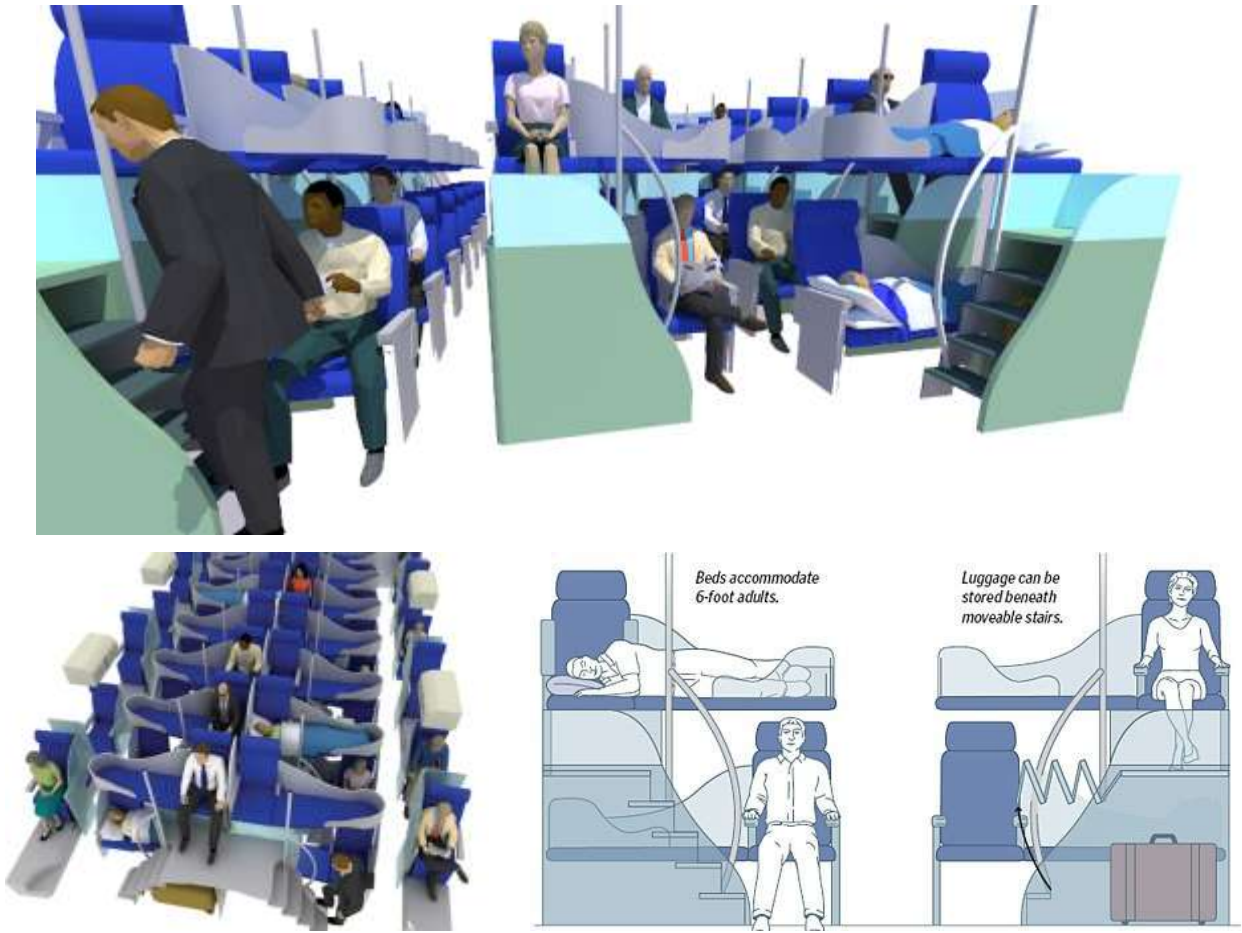

*Source: www.jacob-innovations.com.*

**Figure 9 - "FlexSeat" concept (Jacob-Innovations)**

"FlexSeat" expands the use of vertical space by adopting a two-story logic inspired by large aircraft, buses, and trains, offering superior recline, greater privacy, the possibility of accommodating babies, and additional space for carry-on luggage. The quick conversion between Business and Economy Class creates a capacity management tool that allows supply to be adjusted according to demand, increasing commercial flexibility and the range of fare products. From a segmentation perspective, the airline could explore different levels within each class, distinguishing between upper and lower positions, degrees of privacy, and layouts geared toward families, corporate travelers, or premium passengers on international routes. A more careful analysis, however, points to structural and regulatory challenges, such as safety certification, evacuation, accessibility, and maintenance of a two-level cabin in a reduced space. The heterogeneity of preferences may limit product acceptance, and the operational complexity related to conversion between classes may reduce some of the expected gains, suggesting that the concept tends to be suited to niches or international operations with greater tolerance for innovative solutions and a greater willingness to pay.

Figure 10 . The developer points out that the concept can be adapted to existing Economy or Premium Economy Class areas of aircraft without losing current seats, increasing capacity by 20% on long-haul aircraft.

CAER | Communications in Airline Economics Research, 202117818, 2025.                @ 2025 Center for Airline Economics, Brazil.        15

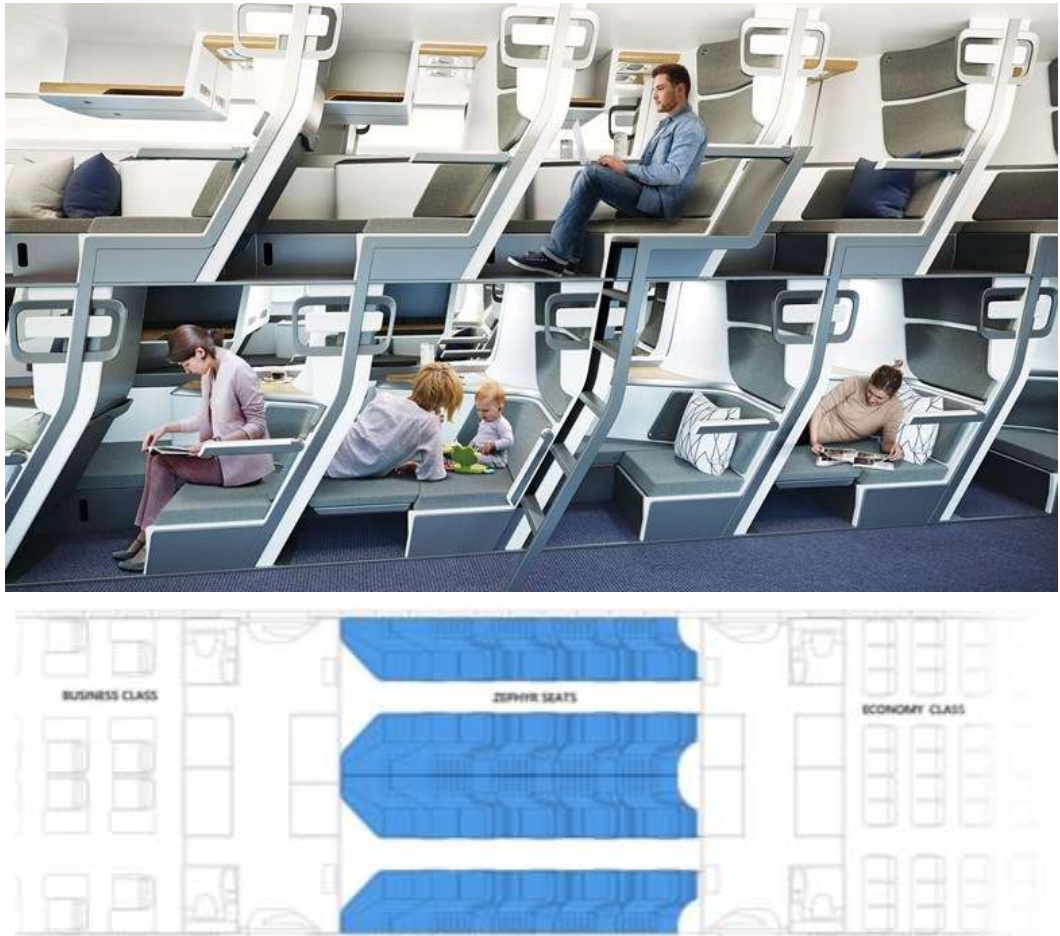

*Sources: www.linkedin.com/company/zephyrseat/about/; www.dailyhive.com/mapped/double-decker-zephyr-seat-airplane; www.republic.co/zephyr-aerospace.*

**Figure 10 - "Zephyr Seat" concept, "Affordable lie-flat airplane seating" (Zephyr Aerospace)**

The "Zephyr Seat" proposes a two-story configuration adaptable to existing Economy or Premium Economy areas, maintaining the current seats and adding additional seats on the upper level, which can increase capacity by about 20 percent on long-haul aircraft. For the airline, this expansion opens space for more precise segmentation, allowing it to offer superior positions as an intermediate product between Premium Economy and Business Class, with fares set according to privacy, recline, and thermal comfort. The structure also enables sales targeted at specific profiles, such as price-sensitive corporate travelers, long-haul passengers seeking horizontal rest, or customers who value relative isolation in dense cabins. A critical analysis, however, points to challenges such as certification requirements, evacuation, structural stability, and public acceptance, especially among passengers who associate the upper level with confinement or greater difficulty of access.

Figure 11 presents Zodiac Seats' "Cabin Hexagon" concept. The idea of hexagons consists of alternating forward- and backward-facing seats to improve cabin space utilization. This unconventional seating arrangement aims to optimize the number of passengers carried on a flight, avoiding side-to-side contact between passengers[12]. On the other hand, it has the potentially undesirable characteristic of reducing privacy due to greater frontal contact.

---

[12] www.dailymail.co.uk/travel/travel_news/article-3154771/Thought-economy-class-couldn-t-worse-horrifying-hexagonal-seating-means-FACE-passengers.html.

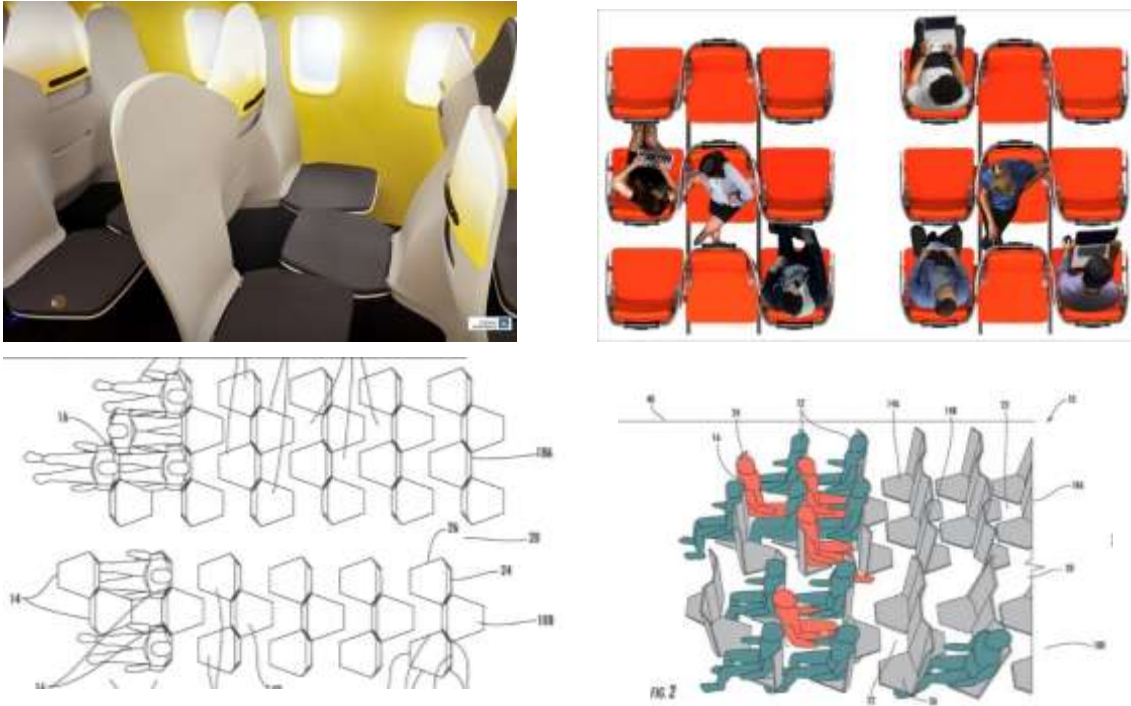

*Sources: www.flyertalk.com/articles/economy-class-cabin-hexagon-promises-more-seats-more-legroom-but-at-will-it-fly.html;*
*www.bluebus.com.br/esse-layout-de-assentos-promete-tornar-as-viagens-de-aviao-ainda-mais-dificeis;*
*www.fortune.com/2015/07/09/new-airplane-seating-hexagon;www.paxex.aero/aviointeriors-janus-economy-class-yin-yang.*

**Figure 11 - "Cabin Hexagon" concept (Zodiac Seats)**

Thus, the "Cabin Hexagon" reorganizes the space by alternating forward- and rear-facing seats, seeking to optimize cabin occupancy and eliminate side-to-side contact between passengers. This geometry allows for increased spatial efficiency and, in some arrangements, increased capacity without compromising individual comfort, creating opportunities for segmentation by position, as certain angles offer an expanded sense of space or better visual access to the aisle. The airline could exploit these differences by charging more for positions in higher demand, offering intermediate categories between standard seats and more comfortable versions, or targeting the product at passengers who are less sensitive to lateral privacy. A critical analysis, however, indicates important limitations. Frontal proximity can reduce privacy, which may hinder acceptance, especially among passengers who avoid prolonged eye contact. There are also issues of certification, ergonomics, and cabin flow management, as well as the possibility of unequal perceptions of comfort between positions in the hexagon.

Figure 12 presents Butterfly's "Checkerboard" concept. It proposes alternating seat configurations on both the x and y axes of the cabin, in a checkerboard pattern. The seats can be folded to give adjacent seats more elbow room—up to two inches more for aisle seats and four inches more for middle seats—plus 20 centimeters (almost 8 inches) of extra legroom for the seats immediately behind[13] . This "checkered board" is a flexible seating system that allows for easy conversion between high-density Economy Class and Business Class on short- and medium-haul flights. The manufacturer emphasizes the "instant conversion" feature, which can be quickly performed by the cabin crew before a flight.

---

[13] www.butterflyseating.com/checkerboard.

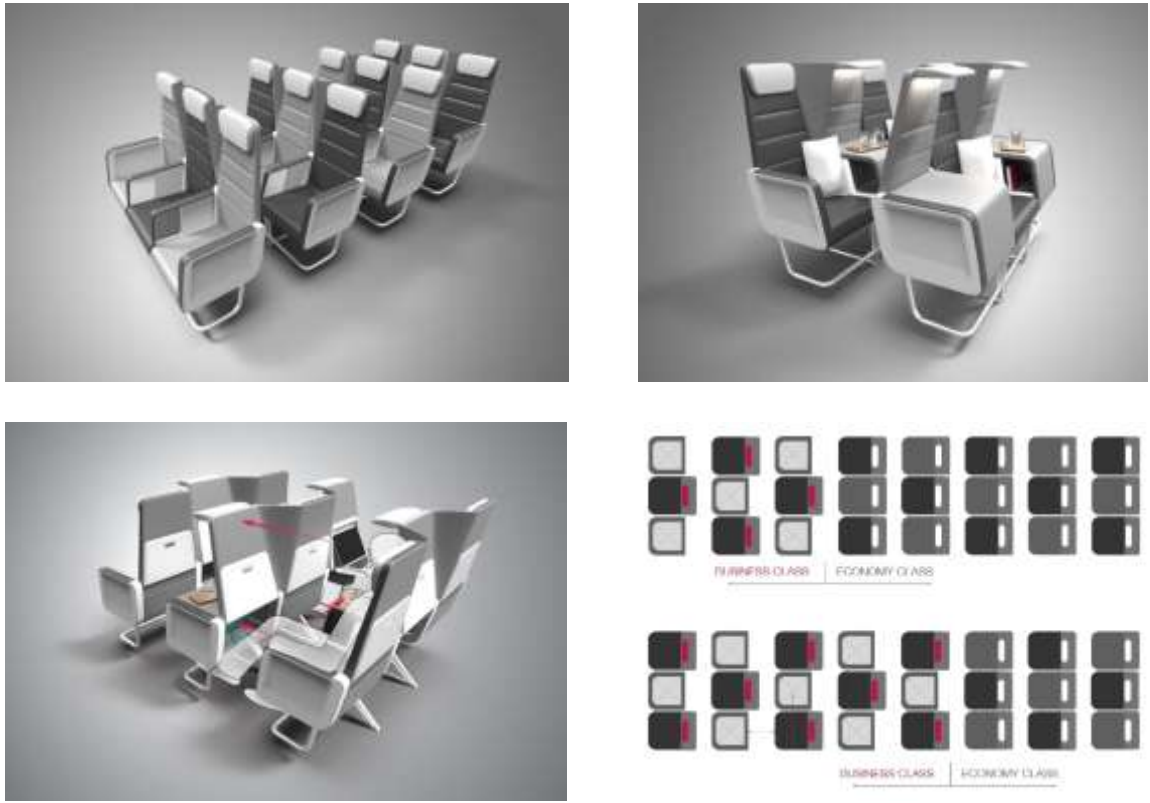

*Source: www.butterflyseating.com/checkerboard.*

**Figure 12 - "Checkerboard Convertible Seating System" (Butterfly) concept**

The "Checkerboard" proposal reorganizes the cabin in a checkerboard pattern on the x and y axes, allowing adjacent seats to be folded down to increase elbow and leg room, creating a geometry that redistributes comfort without reducing the total number of seats. In terms of segmentation, this flexibility opens space for intermediate products within Economy Class, with differentiated fares according to position, range of motion, side privacy, and legroom, as well as enabling compact versions of Business Class on short- and medium-haul routes. The quick conversion between high density and a more comfortable configuration offers companies a tool to adjust supply according to demand, especially in international markets with greater heterogeneity of profiles and greater willingness to pay. A critical analysis, however, indicates limitations such as the risk of perceived inequality between positions, difficulty in managing expectations when additional space depends on folded seats, and challenges in certification, maintenance, and operational standardization. There is also a risk that the experience will be perceived as inconsistent if the configuration varies frequently between flights, which suggests greater suitability for niches or international operations that are more tolerant of differentiated structural solutions.

We conclude our presentation of innovative models with the most controversial of them, the Skyrider, from Aviointeriors (Figure 13). The Skyrider, now in its 3.0 version, is an innovative seat that radically changes the use of cabin space, allowing for ultra-high seat density. Its main feature is its original bicycle saddle-like shape[14] , designed to keep passengers in an upright position in a configuration with a reduced recline and pitch of 23 inches. According to the manufacturer, the design of this seat allows for a 20% increase in the number of passengers on board, in addition to weighing 50% less than conventional standard economy class seats[15] . The Skyrider is controversial because it reinforces the idea of reduced legroom for passengers, which is viewed with suspicion by the industry and the public. However, it cannot fail to be described as innovative, as it presents a new frontier in low-cost travel and potentially opens the possibility of flying to other types of passengers, some of whom currently cannot afford air tickets but who could enter the market if prices were really very low.

---

[14] There are reports that in 2014 Airbus also patented its own bicycle saddle-shaped seat design (source: www.newsweek.com/saddle-seat-economy-flights-could-have-you-standing-la-new-york-888943). It is important to note that the aircraft manufacturing sector is particularly susceptible to the circulation of unverified information and that even established media outlets may, in some cases, reproduce inaccurate or unconfirmed news, as preliminary designs, conceptual studies, and prototypes exhibited at trade shows can generate exaggerated interpretations about their future adoption.
[15] www.aviointeriors.it/2018/press/aviointeriors-skyrider-2-0.

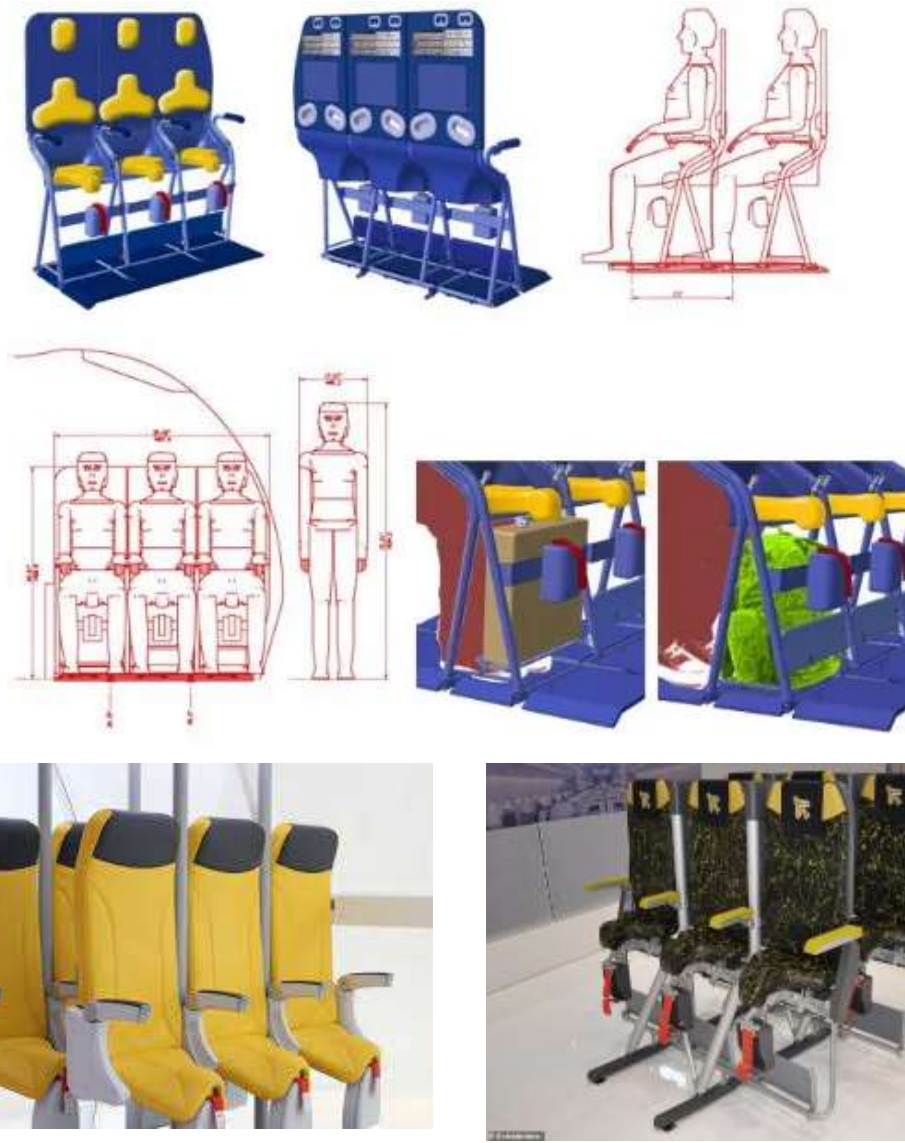

*Sources: www.aviointeriors.it/2018/press/aviointeriors-skyrider-2-0; www.dailymail.co.uk/travel/travel_news/article-6885907/Standing-planes-moves-step-closer-reality-thanks-Aviointeriors-Skyrider-seat.html;*

**Figure 13 - "Skyrider" concept (Aviointeriors)**

The project dates to 2010, when the initial version was presented and promoted as a saddle-type seat designed to reduce the space required for each occupant and, in theory, increase the capacity of single-aisle aircraft, especially on short routes. Starting in 2018, updated versions identified as 2.0 appeared, with structural reinforcement, greater backrest stability, and a pitch of approximately 23 inches. More recent versions, such as 3.0, incorporated simplifications, weight reduction, and minor ergonomic adaptations, maintaining the premise of increasing capacity by about 20 percent and reducing weight by about half compared to traditional seats[16].

The discussion about Skyrider gained visibility when Ryanair expressed interest in exploring ultra-low-cost configurations, mentioning the possibility of vertical or near-standing seats. However, there is no record that any operational testing has been authorized or conducted. Ryanair later denied concrete plans to adopt the concept, although it has used it in its institutional discourse on very low fares and as a strategy to attract engagement.[17]

---

[16] https://www.businessinsider.com/skyrider-standing-airplane-seats-claims-makes-flights-cheaper-2018-4 and various other news outlets.

The feasibility of Skyrider faces significant obstacles. On the regulatory front, near-vertical configurations raise essential certification issues, such as rapid evacuation, protection in turbulence, seat belt suitability, and compliance with minimum ergonomic requirements.[18] On the operational front, its adoption would require a review of boarding procedures, adaptations to cleaning and maintenance logic, and specific crew training to deal with a product that is far from the current standard. In the market, passenger acceptance is uncertain, as perceived discomfort tends to be high even on short flights, and there is a risk of damage to the image of the company that adopts it. The idea of attracting an audience willing to sacrifice extreme comfort in exchange for very low fares may be valid in some markets, but it does not resolve the heterogeneity of preferences and may have negative effects on reputation.

As a result, Skyrider remains an untested innovative concept, whose widespread adoption faces regulatory, operational, and commercial barriers. The proposal illustrates extreme limits of cabin density and ultra-low-cost strategies, but its actual implementation, especially in aircraft such as the Boeing 737-800 operated by Ryanair, seems very unlikely in the short and medium term.

Joint consideration of the innovative concepts illustrated above, as well as the results of the literature in the field (Section II), allows us to construct a broader comparative framework on how seat density, comfort, and operating costs interact in commercial aviation product design. Configurations that increase density, such as Skyrider, Cabin Hexagon, or double-decker economy proposals, follow the logic of reducing unit costs by increasing capacity. Solutions such as Morph Seating, Cozy Suite, or Zephyr Seat work in the opposite direction, increasing comfort and creating micro-products capable of sustaining fare differentiation in competitive markets, in line with evidence that segmented improvements can generate premiums or strengthen revenues when the public is less price sensitive. In both cases, the link between cabin structure, differentiated elasticities, and pricing strategies remains. Thus, both concepts that increase density and those that enhance comfort show that the internal layout is an economic tool that shapes operating costs, perceived quality, and forms of fare segmentation.

The Skyrider tends to be less viable than the other concepts analyzed. Although several projects face challenges of certification, acceptance, and operational complexity, the Skyrider positions this set of difficulties at a more critical level, despite being probably the one that attracts the most public attention. The reason for its unfeasibility is that the concept changes not only the geometry of the cabin but the very definition of a seat, bringing the experience closer to an almost vertical posture with saddle-type support. As discussed above, this contradicts safety standards, minimum ergonomics, evacuation requirements, and impact protection, creating more rigid barriers than those found in double-decker, pod, or staggered seat arrangements. In addition, other designs offer some noticeable gains in comfort, privacy, or modularity, even when associated with higher density. The Skyrider, on the other hand, emphasizes extreme capacity increases with an explicit loss of comfort, which makes it difficult to reconcile with passenger perceptions of value and the airline's image. This asymmetry limits acceptance, reduces willingness to pay, and restricts the target audience to passengers who are extremely price-sensitive on short routes.

1 presents a comparative summary of the innovative cabin layout concepts discussed in this section, organizing information about the manufacturer, approximate date of market introduction, most suitable segment, and potential effects on seat density and commercial segmentation.

---

[17] https://www.euronews.com/business/2025/05/23/ryanair-denies-claims-flights-will-soon-offer-cheaper-standing-seats.
[18] https://www.airlineratings.com/articles/we-call-bs-on-these-airline-seats.

**Table 1 - Comparison of the main innovative cabin layout concepts according to density and segmentation**

| Concept | | Manufacturer | Approximate launch year | Most appropriate segment | Potential effect on density | Potential effect on segmentation |
|---|---|---|---|---|---|---|
| Morph Seating | 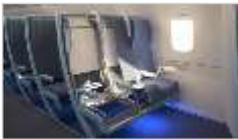 | Seymourpowell | 2013 | Domestic or International | Medium | Medium |
| Cozy Suite | 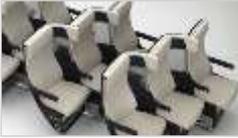 | Thompson Aero Seating | 2008 | Domestic or International | Medium | Medium |
| Air Lair | 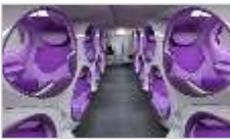 | Factorydesign | 2012 | International (long haul) | High | High |
| StepSeat | 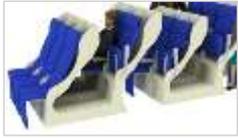 | Jacob-Innovations | 2014 | Domestic or International (short and medium routes) | Low | Low |
| FlexSeat | 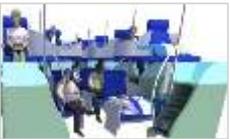 | Jacob-Innovations | 2006 | International (long haul) | High | High |
| Zephyr Seat | 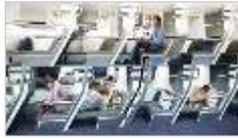 | Zephyr Aerospace | 2020 | International (long haul) | High | High |
| Cabin Hexagon | 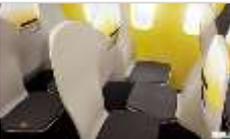 | Zodiac Seats (Zodiac Aerospace France) | 2015 | Domestic or International | Medium | Medium |
| Checkerboard | 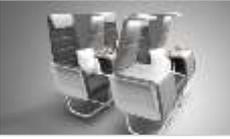 | Butterfly / Paperclip Design | 2012 | Domestic or International | Low | Medium |
| Skyrider | 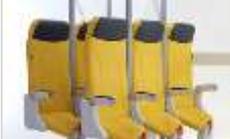 | Aviointeriors | 2010 | Domestic (short routes) | High | Low |

*Sources: Various websites specializing in commercial aviation, manufacturers' institutional pages, online industry reports, Wikipedia. Authors' own qualitative classifications.*

In Table 1 , the ratings of potential effects on density and segmentation (last columns on the right) are from the authors, based on an analysis of the structural characteristics of each concept and the associated operating costs, as discussed in various technical and industry sources. Concepts classified as having a high impact on density, such as Air Lair, Zephyr Seat, and Skyrider, share the intensive use of vertical space or the reduction of individual space, even with radically opposite strategies, but allowing for the addition of new levels or the compression of seats to the point of creating significant additional capacity. Medium-impact concepts, such as Morph Seating, Cozy Suite, FlexSeat, and Cabin Hexagon, promote structural rearrangements that improve spatial efficiency or introduce new geometries, but without profoundly altering the internal architecture of the cabin. Finally, StepSeat and Checkerboard remain low impact, as they preserve the total number of seats, acting mainly on the redistribution of comfort, ergonomics, or lateral privacy. The comparative classification allows us to identify which proposals are geared toward maximizing capacity and which favor product differentiation aimed at fare segmentation.

### III.3. EXPLORATORY MAPPING OF ALTERNATIVE CABIN CONFIGURATIONS

Based on the set of innovative seating concepts presented above, this section aims to illustrate, through an example, some possibilities for cabin interior arrangements in the context of Brazilian domestic commercial aviation. The aim is to show, in a visual and direct way, how changes in pitch, economy class distribution, and number of rows can modify the density and spatial organization of the cabin. This is a descriptive exercise that helps to understand how structural variations in layout can interact, in later stages, with issues such as comfort, product differentiation, and fare structure.

14 presents three alternative configurations of a Boeing 737-800 used in domestic operations. In the figure, the notation (P, R, S) indicates, for a given area of the cabin, respectively, the seat pitch in inches P, the number of rows of seats R, and the total number of seats S. Thus, for example, (P, R, S) equal to (31, 31, 183) should be read as 31-inch pitch, 31 rows, and 183 available seats. The first two configurations, shown at the top of the figure, were actually used by Gol in 2014 and have capacities for 183 and 177 passengers. In both cases, the main pitch is 31 inches. In the 177-seat configuration, there are two economy classes, called "CE/C" (Comfort Economy Class) and "SE/C" (Standard Economy Class). In CE/C, the pitch is 34 inches, covering 42 seats, i.e., (P, R, S) equal to (34, 7, 42).

The third configuration, shown at the bottom of 14 , is fictitious and serves only to illustrate ideas about the role of row density inside the cabin. This is an illustrative representation, entirely unrelated to any actual analysis of weight, balance, certification, or aircraft manufacturing engineering, which is why we strongly suggest its use only as a visual reference, with future technical feasibility studies being necessary. In this configuration, CE/C is maintained with a pitch of 34 inches, SE/C is preserved with a pitch of 31 inches, and a third economy class, "BE/C" (Basic Economy Class), is added, based on the Skyrider seat concept (vertical, saddle-shaped), represented in purple. In this class, the pitch would be 23 inches, as indicated by the developer. The inclusion of this class would increase the total capacity to 195 passengers. Even after discussing the challenges and feasibility of this type of seat row configuration in modern commercial aviation, Figure 14 allows us to immediately visualize how different layout proposals alter density, occupancy, and internal distribution, helping us to understand the problem and interpret the relationships between capacity, comfort, and cabin spatial organization.

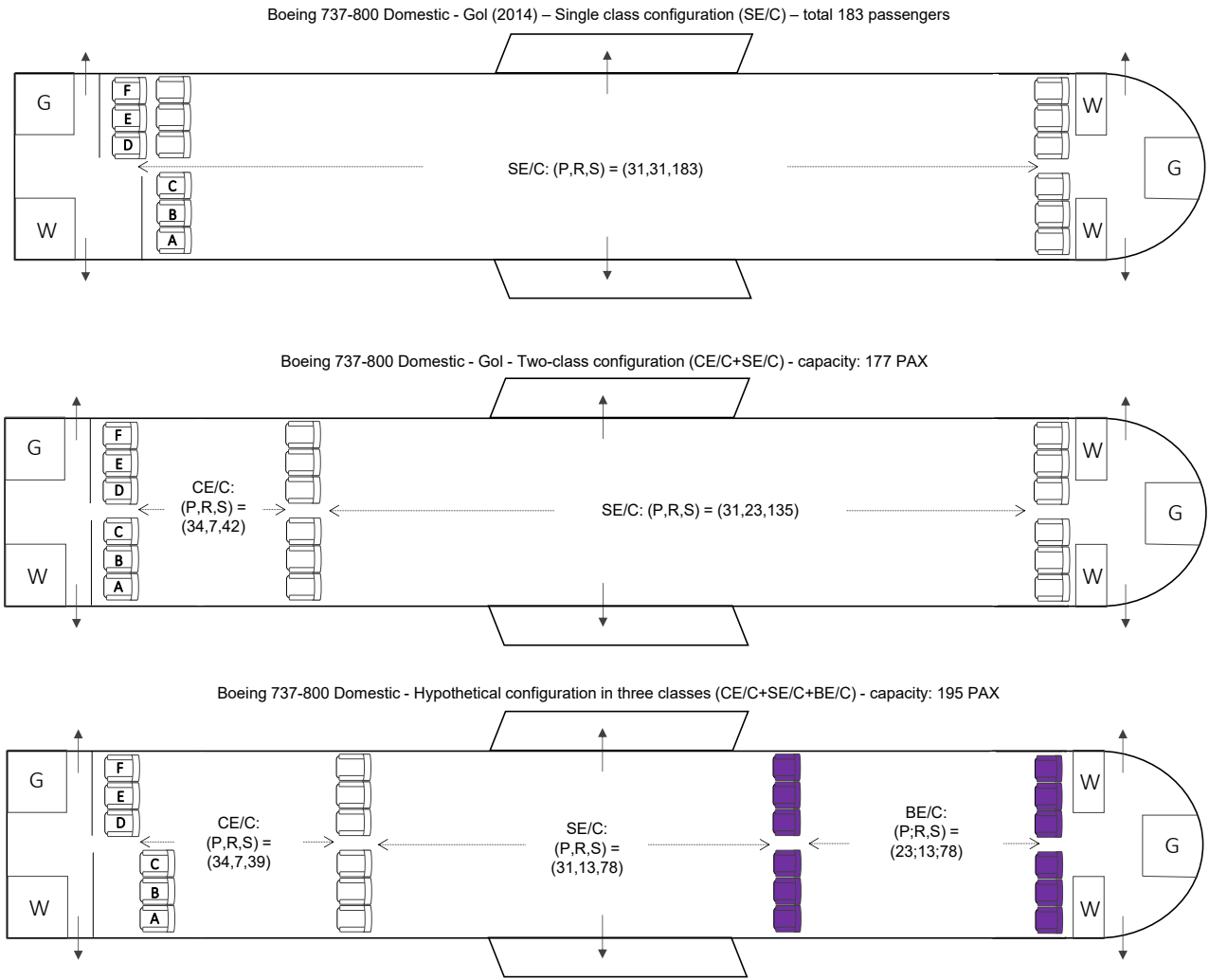

*Note: SE/C = "Standard Economy" class; CE/C = "Comfort Economy" class; BE/C = "Basic Economy" class; P, R, and S denote, respectively, seat pitch, number of seat rows, and total seats in the cabin.*

**Figure 14 - Aircraft configurations - Gol and fictional example of the introduction of an innovative layout concept**

# IV. CONCEPTUAL MODEL

Studying the market effects of cabin layout decisions requires a broad understanding of the competitive environment of air transport. These decisions involve, on the one hand, cabin density, i.e., how many rows and how much space between seats will be offered, and, on the other hand, passenger segmentation, defined by whether to create sub-products within economy class, such as premium seats or differentiated areas on the aircraft. These choices are linked to travel demand, airfare pricing strategies, ancillary revenue generation, and the operating cost structure of companies. Together, these elements shape the competitive positioning of each airline and directly influence the profile of passengers it can attract and retain. Given the complexity of the problem, this section aims to propose a conceptual framework that serves as a basis and guide for the empirical analyses to be developed.

Figure 15 presents a decision tree diagram, structured as a sequential game, which allows us to visualize the decision-making process of airlines regarding the internal configuration of their aircraft cabins, as well as the implications that these decisions have for the configuration of their business model. We call the diagram "Decision game for aircraft cabin configuration and airline market positioning," which serves to illustrate the strategic interaction involved and guides our analysis. Note that this decision-making environment is surrounded by global market competition among companies, which involves consideration not only of their own business model, but also of the business models of rival companies, in a process of strategic interdependence. As can be seen in Figure 15, a given airline ("Airline 1") must decide whether to "segment" or "not segment" its aircraft cabin by introducing one or more premium economy classes in addition to the traditional economy class. Simultaneously with this decision, the airline must define the degree of cabin densification, adding more or fewer rows of seats, either by specifying new orders to the manufacturer or by adapting used aircraft incorporated into the fleet during the "fleet rollover" process (gradual replacement of the fleet with newer or different aircraft). It must therefore decide between "densify" or "not densify." In turn, a rival airline ("Airline 2") must make similar decisions, considering the expected decision-making of the competition. With other companies in the industry, each is expected to position itself strategically in the market to achieve a balance with companies focused on the "Ultra Low Cost," "Low Cost," and "Mainline" (traditional) business models.

The decisions made by each airline throughout the game will define its price positioning with consumers, obtaining a larger or smaller market share among passengers who are more or less price sensitive ("Low-yield PAX" and "High-yield PAX"). This positioning will be decisive in shaping its business model. Thus, the final branches of the tree represent four possible strategic positions for airlines, each resulting from the combination of whether or not to segment the cabin and whether or not to increase seat density. For simplicity, the diagram at Figure 15 shows the results for Airline 2, conditional on Airline 1 opting for cabin densification, but other results are possible. Thus, each result corresponds to a distinct type of market performance:

- When the company does not segment and increases cabin density, the result is a positioning focused on low-yield passengers (low price per kilometer), i.e., those who are very price sensitive. This combination leads to a reduced cost structure and tends to characterize Ultra Low-Cost models, whose focus is to maximize capacity and offer minimum fares.
- When the company does not segment and does not increase density, it operates with lower density and without creating sub-products within the economy class. This leads to a greater focus on high-yield passengers (high price per kilometer), but those who are less price-sensitive and who value relative comfort more. This is a typical positioning for traditional Mainline companies.
- When the company segments and densifies, the result is a mix of low-yield and high-yield passengers, as the presence of a premium product attracts consumers who are more willing to pay, while densification keeps costs low and allows price-oriented customers to be captured. This arrangement can be found in both low-cost companies and mainline companies with hybrid strategies.

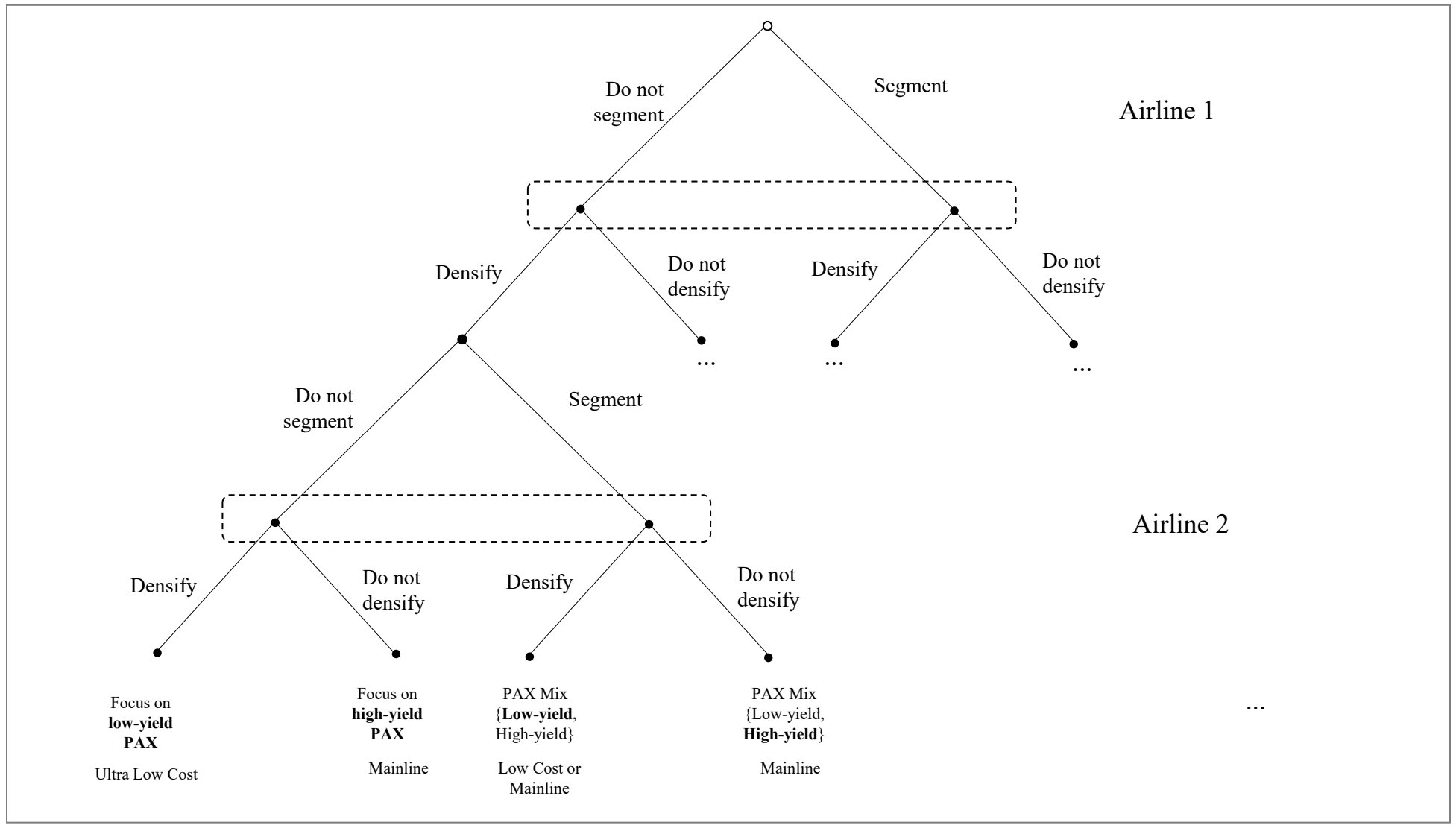

*Note: The diagram illustrates, in a simplified way, a sequential game between two fictional airlines in choosing between "Segment" or "Do Not Segment" and "Dense" or "Do Not Dense" the cabin, considering aircraft equipped with Economy Class only. The final branches illustrate the final results for Airline 2 conditional on Airline 1's choice to densify, although other possible outcomes are not represented. This structure summarizes the strategic interaction involved and the implications of these decisions for the companies' market positioning and the configuration of their business model, which can be "Ultra Low Cost," "Low Cost," and "Mainline" (traditional). "Low-yield PAX" are passengers who are more price sensitive, while "High-yield PAX" are passengers who are less price sensitive and associated with higher fares.*

**Figure 15 - Decision-making game for aircraft cabin configuration and market positioning of airlines**

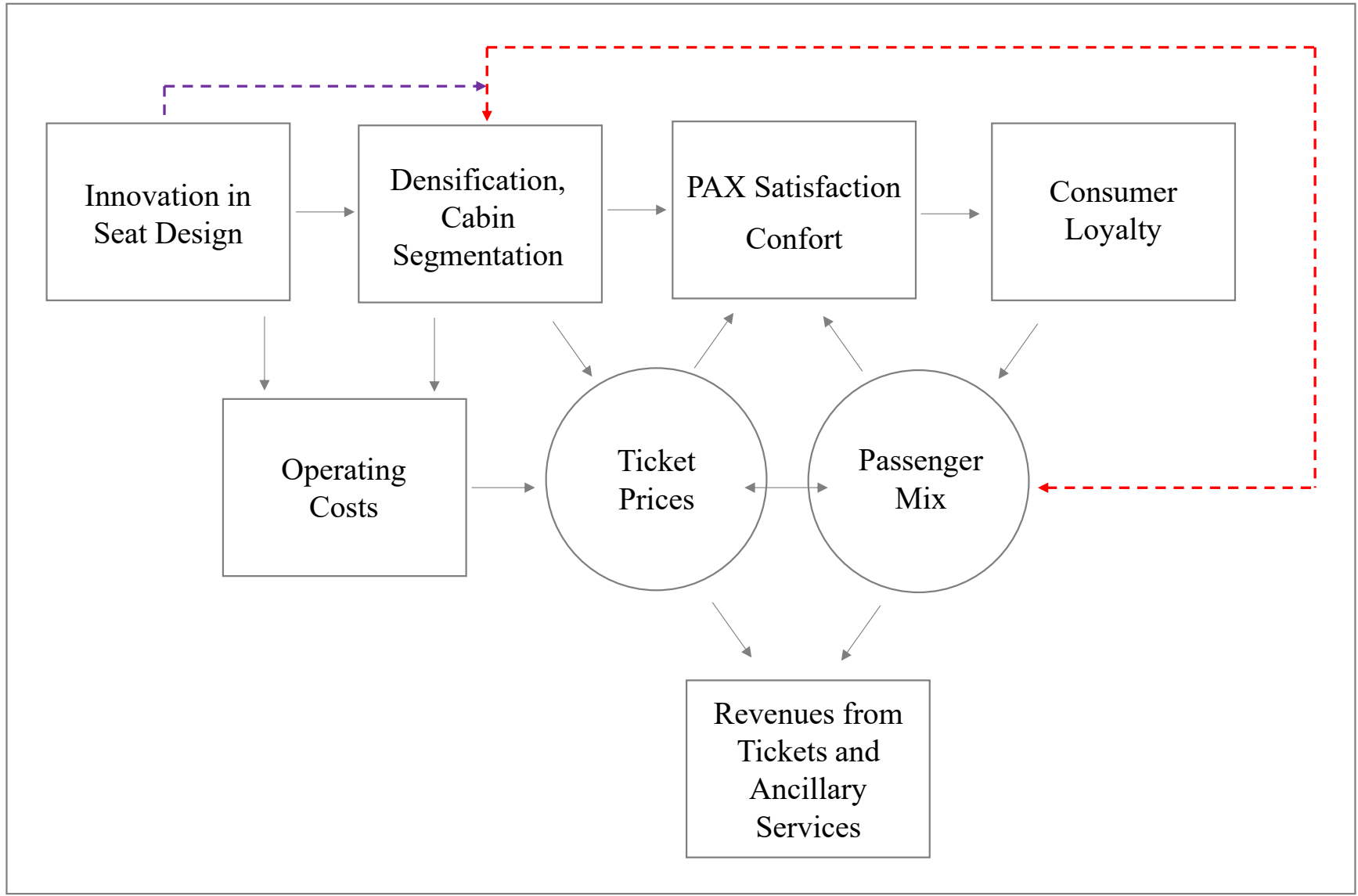

**Figure 16 - Conceptual model of the impact of introducing innovative seat design**

- When the company segments rather than densifies, it offers relatively more space in addition to differentiated economy classes. This result also generates a mix of passengers, but with greater emphasis on high-yield consumers who value comfort and additional services. This positioning tends to be associated with Mainline companies to capture higher-value passengers.

Together, these four outcomes of the cabin configuration game suggest how choices about cabin segmentation and density define the audience served, the business model, and the competitive positioning of each company in the airline industry.

Figure 16 presents the conceptual model proposed to analyze the market impacts of decisions regarding seat density in aircraft and the introduction of innovative seat design in Brazilian commercial aviation. The starting point for the model is the introduction of an "Innovation in Seat Design," treated as an external shock that directly alters two elements of supply: "Densification, Cabin Segmentation" and "Operating Costs." Densification refers to the increase or reduction in the number of rows and the space between seats, while segmentation refers to the creation of sub-products within economy class, such as premium seats, as discussed above. At the same time, changes in these two elements affect price behavior in the market, represented by the "Ticket Prices" block. These prices are determined by costs, consumers' willingness to pay, and the competitive positioning of companies. Innovation also influences "Passenger Satisfaction and Comfort," which in turn directly affects "Consumer Loyalty," an essential element for retaining frequent flyers and ensuring revenue sustainability over time.

At the center of the conceptual model of Figure 16 are the two circular blocks, "Ticket Prices" and "Passenger Mix," whose interaction is endogenous. Changes in price attract different passenger profiles, modifying the mix between consumers who are more price sensitive and those willing to pay more for comfort or additional services. This composition directly affects the total revenue obtained from tickets and extra fees, grouped in the "Revenue from Tickets and Ancillary Services" block. The inverse relationship also holds: changes in the "Passenger Mix" put pressure on the price structure, requiring adjustments to balance costs, demand, and competitiveness. Thus, we see that the model presented in Figure 16 shows how an innovation initially introduced inside the cabin triggers a sequence of effects on cost structure, pricing, demand composition, passenger satisfaction, and revenue generation, creating a dynamic competitive environment in which small design changes can have significant systemic impacts.

The complexity of the strategic game and conceptual model presented shows that cabin configuration decisions involve multiple simultaneous interactions between costs, prices, passenger satisfaction, ancillary revenues, loyalty, and demand composition. Each element feeds back into the others, forming a system in which small changes, such as a change in seat design, can generate amplified effects across the entire competitive structure of the sector. Given this intrinsically interdependent environment, this study focuses only on a specific part of this set of relationships, focusing on the impact of layout choices on "Ticket Prices." This analytical decision recognizes that prices are endogenous to the "Passenger Mix," but, at this early stage, we treat this relationship as exogenous to enable empirical identification. A full assessment of the endogeneity between these two blocks is left for future studies, which may explore in greater depth the feedback mechanisms between demand profile and pricing strategies.

By adopting this more restricted approach, we start from the premise that innovations in seat design are directly reflected in decisions on "Cabin Densification and Segmentation," as observed in the study by Lee & Luengo Prado (2004), which analyzed the effects of pitch changes and the introduction of differentiated classes in the early 2000s. A similar situation occurs with the phenomenon we are investigating, related to the comfort seats introduced in the mid-2010s, whose most immediate impacts appear in the price differences between products within the same aircraft. Thus, the econometric model developed here is embedded in the logic of the conceptual framework presented, but represents only a small part of a broader and more complex reality. This reinforces the need for cautious interpretations and future extensions that can more fully capture the dynamic structure that characterizes the commercial aviation market.

# V. DATA

Most of the data used in this study comes from a survey of passengers at Brazilian airports, commissioned by the Planning and Logistics Company (EPL) in 2014. The EPL report (2014) presents the official results of this initiative, called the "Air Passenger Transport Origin/Destination Survey." The main objective of the survey was to expand knowledge about the flow of people using air transport in Brazil, both for domestic and international travel, by building a comprehensive panel of travelers' behavior and characteristics. The survey was conducted by the Olhar Institute, Research and Strategic Information, responsible for the collection and initial systematization of the information.

The EPL survey was designed to support the formulation of the National Integrated Logistics Plan (PNLI), an instrument aimed at proposing actions for the development and integration of the country's various modes of transport. According to EPL (2014), the sample design covered 65 airports, representing 53.7% of airports with regular flights and, simultaneously, 99% of all passengers boarding in Brazil two years prior to data collection. The questionnaires were administered in the domestic and international departure areas of the selected airports, in four stages throughout 2014 (January/February, March/April, May, and August), as described in the study's methodological procedures.

For this research, the original EPL database was supplemented with a set of additional information. To enable more detailed analyses of the location of passengers on board, tens of thousands of boarding passes previously scanned in image format for research audit purposes were subjected to a process of transcription and tabulation of data relating to passenger seat designators (such as seats with codes 5A, 14F, or 22C), structuring them in a format suitable for analysis. This process made it possible to identify the exact location of the seats occupied by passengers and cross-reference these records with the information collected in the questionnaire interviews.

In addition, several other databases from the National Civil Aviation Agency (ANAC) were integrated into the study, providing greater depth and accuracy to the analyses. Among them are: the Air Transport Statistical Database, with aggregate data and microdata; Microdata on Commercial Air Fares; the Active Regular Flight (VRA) database, which compiles detailed flight history; the Brazilian Aeronautical Registry (RAB), with information on the national fleet; and the Operations Registration System (SIROS). The combination of these multiple sources made it possible to consolidate a detailed database to meet the needs of the proposed investigations.

2 presents a descriptive analysis of the available data. It contains a passenger dispersion map (number and percentages of respondents) according to seat location on the aircraft. The table shows the distribution of passengers interviewed by row and seat, allowing us to observe how travelers are dispersed throughout the cabin. The left panel shows the absolute numbers, the center panel expresses these same quantities as percentages per row, and the right panel shows the relative share of each seat in the total sample. There is a consistent pattern of higher occupancy of aisle seats, especially seat D, which accounts for more than 25 percent in several rows, followed by seats C and E. Window seats, A and F, have an intermediate share, while the middle seat, B, has a smaller share in most rows, reinforcing the perception of reduced preference for this position. There is also a slight gradient toward the center of the cabin, where there is a higher concentration of respondents, suggesting that for aircraft with rear and front boarding, occupancy is distributed relatively evenly. These patterns help to understand how passengers' structural preferences can influence pricing and demand for certain types of seats.

**Table 2 - Passenger dispersion map (number and percentages of respondents), according to seat location**

| | | Seat Positions | | | | | | | | | Seat Positions | | | | | | | | | Seat Positions | | | | | |
|---|---|---|---|---|---|---|---|---|---|---|---|---|---|---|---|---|---|---|---|---|---|---|---|---|---|
| | | A | B | C | D | E | F | Total | | | A | B | C | D | E | F | Total | | | A | B | C | D | E | F | Total |
| | 1 | 224 | 130 | 215 | 374 | 71 | 179 | 1,193 | | 1 | 19% | 11% | 18% | 31% | 6% | 15% | 100% | | 1 | 1% | 2% | 2% | 3% | 2% | 2% | 2% |
| | 2 | 527 | 321 | 388 | 550 | 168 | 331 | 2,285 | | 2 | 23% | 14% | 17% | 24% | 7% | 14% | 100% | | 2 | 3% | 4% | 3% | 4% | 4% | 3% | 4% |
| | 3 | 731 | 364 | 496 | 614 | 206 | 388 | 2,799 | | 3 | 26% | 13% | 18% | 22% | 7% | 14% | 100% | | 3 | 4% | 5% | 4% | 4% | 5% | 4% | 4% |
| | 4 | 712 | 364 | 487 | 660 | 216 | 426 | 2,865 | | 4 | 25% | 13% | 17% | 23% | 8% | 15% | 100% | | 4 | 4% | 5% | 4% | 5% | 5% | 4% | 4% |
| | 5 | 723 | 338 | 525 | 640 | 200 | 379 | 2,805 | | 5 | 26% | 12% | 19% | 23% | 7% | 14% | 100% | | 5 | 4% | 4% | 5% | 4% | 5% | 4% | 4% |
| | 6 | 717 | 436 | 611 | 698 | 192 | 408 | 3,062 | | 6 | 23% | 14% | 20% | 23% | 6% | 13% | 100% | | 6 | 4% | 5% | 5% | 5% | 5% | 4% | 5% |
| | 7 | 704 | 411 | 599 | 703 | 200 | 387 | 3,004 | | 7 | 23% | 14% | 20% | 23% | 7% | 13% | 100% | | 7 | 4% | 5% | 5% | 5% | 5% | 4% | 5% |
| | 8 | 694 | 406 | 606 | 694 | 173 | 371 | 2,944 | | 8 | 24% | 14% | 21% | 24% | 6% | 13% | 100% | | 8 | 4% | 5% | 5% | 5% | 4% | 4% | 5% |
| | 9 | 696 | 395 | 546 | 684 | 154 | 368 | 2,843 | | 9 | 24% | 14% | 19% | 24% | 5% | 13% | 100% | | 9 | 4% | 5% | 5% | 5% | 4% | 4% | 4% |
| | 10 | 692 | 356 | 566 | 624 | 168 | 363 | 2,769 | | 10 | 25% | 13% | 20% | 23% | 6% | 13% | 100% | | 10 | 4% | 4% | 5% | 4% | 4% | 4% | 4% |
| | 11 | 621 | 288 | 420 | 535 | 123 | 282 | 2,269 | | 11 | 27% | 13% | 19% | 24% | 5% | 12% | 100% | | 11 | 4% | 4% | 4% | 4% | 3% | 3% | 4% |
| | 12 | 634 | 303 | 388 | 594 | 126 | 302 | 2,347 | | 12 | 27% | 13% | 17% | 25% | 5% | 13% | 100% | | 12 | 4% | 4% | 3% | 4% | 3% | 3% | 4% |
| Seating Rows | 13 | 450 | 215 | 251 | 389 | 55 | 157 | 1,517 | Seating Rows | 13 | 30% | 14% | 17% | 26% | 4% | 10% | 100% | Seating Rows | 13 | 3% | 3% | 2% | 3% | 1% | 2% | 2% |
| | 14 | 582 | 328 | 399 | 538 | 144 | 316 | 2,307 | | 14 | 25% | 14% | 17% | 23% | 6% | 14% | 100% | | 14 | 3% | 4% | 3% | 4% | 3% | 3% | 4% |
| | 15 | 665 | 348 | 418 | 584 | 119 | 282 | 2,416 | | 15 | 28% | 14% | 17% | 24% | 5% | 12% | 100% | | 15 | 4% | 4% | 4% | 4% | 3% | 3% | 4% |
| | 16 | 578 | 311 | 381 | 520 | 86 | 269 | 2,145 | | 16 | 27% | 14% | 18% | 24% | 4% | 13% | 100% | | 16 | 3% | 4% | 3% | 4% | 2% | 3% | 3% |
| | 17 | 605 | 278 | 417 | 506 | 157 | 339 | 2,302 | | 17 | 26% | 12% | 18% | 22% | 7% | 15% | 100% | | 17 | 4% | 3% | 4% | 3% | 4% | 3% | 4% |
| | 18 | 580 | 246 | 369 | 468 | 157 | 372 | 2,192 | | 18 | 26% | 11% | 17% | 21% | 7% | 17% | 100% | | 18 | 3% | 3% | 3% | 3% | 4% | 4% | 3% |
| | 19 | 535 | 226 | 370 | 458 | 170 | 431 | 2,190 | | 19 | 24% | 10% | 17% | 21% | 8% | 20% | 100% | | 19 | 3% | 3% | 3% | 3% | 4% | 4% | 3% |
| | 20 | 560 | 224 | 380 | 418 | 178 | 437 | 2,197 | | 20 | 25% | 10% | 17% | 19% | 8% | 20% | 100% | | 20 | 3% | 3% | 3% | 3% | 4% | 4% | 3% |
| | 21 | 610 | 231 | 352 | 416 | 161 | 384 | 2,154 | | 21 | 28% | 11% | 16% | 19% | 7% | 18% | 100% | | 21 | 4% | 3% | 3% | 3% | 4% | 4% | 3% |
| | 22 | 550 | 213 | 351 | 433 | 155 | 381 | 2,083 | | 22 | 26% | 10% | 17% | 21% | 7% | 18% | 100% | | 22 | 3% | 3% | 3% | 3% | 4% | 4% | 3% |
| | 23 | 574 | 217 | 355 | 429 | 171 | 402 | 2,148 | | 23 | 27% | 10% | 17% | 20% | 8% | 19% | 100% | | 23 | 3% | 3% | 3% | 3% | 4% | 4% | 3% |
| | 24 | 433 | 203 | 308 | 388 | 121 | 318 | 1,771 | | 24 | 24% | 11% | 17% | 22% | 7% | 18% | 100% | | 24 | 3% | 3% | 3% | 3% | 3% | 3% | 3% |
| | 25 | 420 | 166 | 215 | 329 | 81 | 260 | 1,471 | | 25 | 29% | 11% | 15% | 22% | 6% | 18% | 100% | | 25 | 3% | 2% | 2% | 2% | 2% | 3% | 2% |
| | 26 | 385 | 167 | 226 | 296 | 96 | 261 | 1,431 | | 26 | 27% | 12% | 16% | 21% | 7% | 18% | 100% | | 26 | 2% | 2% | 2% | 2% | 2% | 3% | 2% |
| | 27 | 385 | 157 | 224 | 291 | 109 | 262 | 1,428 | | 27 | 27% | 11% | 16% | 20% | 8% | 18% | 100% | | 27 | 2% | 2% | 2% | 2% | 3% | 3% | 2% |
| | 28 | 360 | 131 | 176 | 249 | 76 | 251 | 1,243 | | 28 | 29% | 11% | 14% | 20% | 6% | 20% | 100% | | 28 | 2% | 2% | 2% | 2% | 2% | 3% | 2% |
| | 29 | 289 | 108 | 168 | 230 | 75 | 216 | 1,086 | | 29 | 27% | 10% | 15% | 21% | 7% | 20% | 100% | | 29 | 2% | 1% | 1% | 2% | 2% | 2% | 2% |
| | 30 | 186 | 65 | 126 | 146 | 51 | 144 | 718 | | 30 | 26% | 9% | 18% | 20% | 7% | 20% | 100% | | 30 | 1% | 1% | 1% | 1% | 1% | 1% | 1% |
| | 31 | 125 | 34 | 82 | 75 | 34 | 113 | 463 | | 31 | 27% | 7% | 18% | 16% | 7% | 24% | 100% | | 31 | 1% | 0% | 1% | 1% | 1% | 1% | 1% |
| | 32 | 85 | 31 | 54 | 40 | 18 | 93 | 321 | | 32 | 26% | 10% | 17% | 12% | 6% | 29% | 100% | | 32 | 1% | 0% | 0% | 0% | 0% | 1% | 0% |
| | Total | 16,632 | 8,011 | 11,469 | 14,573 | 4,211 | 9,872 | 64,768 | | Total | 26% | 12% | 18% | 23% | 7% | 15% | 100% | | Total | 100% | 100% | 100% | 100% | 100% | 100% | 100% |

*Notes: Analysis of data from the Passenger Air Transport Origin/Destination Survey (EPL, 2014), compiled version. The table does not represent the configuration of an aircraft, but only allows the visualization of spatial dispersion, considering different types of aircraft and configurations. The colors represent the relative intensity of occupancy: shades of blue indicate lower concentrations of passengers and shades of red indicate higher concentrations, facilitating the visual identification of seat preference patterns and distribution across rows. Of the 122,000 passengers interviewed, only those with domestic flights from the four largest airlines with identified flights were selected. Questionnaires with errors or omissions, photographs of illegible or cut boarding passes, and rows above number 32 were disregarded.*

# VI. METHODOLOGY

This study aims to estimate the effect of seat density, as well as seat location within the cabin layout, on airfare prices. The empirical motivation is based on evidence that preferences related to space and seat selection influence pricing patterns. As seen, Rouncivell, Timmis, and Ison (2018) show that price sensitivity has a negative relationship with the willingness to pay for seat selection, indicating that purchasing decisions incorporate trade-offs between cost and convenience. Lee and Luengo Prado (2004) document that adjustments in pitch affect ticket prices, albeit differently between strategies of uniform increase or increase restricted to specific rows. These previous studies illustrate how physical attributes of the cabin and differentiation policies can produce systematic variations in observed prices, a rationale that guides the empirical specification adopted in this analysis.

As seen, the database consists of the EPL Survey, Passenger Profile Study, with face-to-face interviews at Brazilian airports and collection of boarding passes. The sample used in the estimates contains 15,517 observations after cleaning and matching (in some simpler model specifications, 15,634), concentrated on flights operated by Gol and TAM (now Latam).

The set of variables used in the econometric modeling is presented below. For estimation purposes, all continuous variables were transformed into logarithms.

- P is the dependent variable of the models, being the price paid for the airline ticket, as stated by the passenger in the EPL Survey questionnaires (source: EPL Survey questionnaires). The variable is expressed in current monetary units (R$).
- ADV is the advance purchase in days, reported by the passenger in the EPL Survey (source: EPL Survey questionnaires). The variable measures how many days passed between the purchase and the flight, allowing us to observe how the time of purchase relates to the price.
- DIST is the distance between airports on the route, in kilometers, obtained from official statistical records (source: ANAC Air Transport Statistical Data, own calculations). The variable is linked to flight costs and possible competition with ground transportation modes.
- BSN is the indication of business travel, declared directly by the passenger in the EPL Survey (source: EPL Survey questionnaires). The variable distinguishes traveler profiles and aims to capture possible differences in price sensitivity that manifest themselves via the price paid for tickets.
- FLTIME is the flight time in minutes, obtained from official operational data (source: ANAC Air Transport Statistics, own calculations). The variable is associated with flight costs, competition with ground transportation, and discomfort or time costs during the trip.
- SHIPMENT is the total volume transported in the hold, in kilograms, including baggage and cargo (source: ANAC Air Transport Statistics, own calculations). The variable represents the use of the hold and is related to costs and processing time linked to the aircraft's turnaround time.
- REVPAX is the number of paying passengers on board, obtained from official statistical data (source: ANAC Air Transport Statistical Data, own calculations). The variable indicates the traffic density of the route and its operational effects.
- LF is the load factor, defined as the percentage of seats offered on the aircraft that were occupied by paying passengers based on official records (source: ANAC Air Transport Statistics, own calculations). The variable shows the level of occupancy and seat scarcity.
- FUELP is the price of fuel in the region of the airports on the route, expressed in current monetary units (R$), obtained from consolidated records (source: National Petroleum Agency, ANP, own calculations). The variable represents the cost component associated with the energy input used in the flight.
- HUB is a dummy variable indicating that the passenger made at least one connection, as reported in the EPL Survey (source: EPL Survey questionnaires). The variable reflects network conditions and airport accessibility.
- SEATSH is the airline's share of the total seats offered on the route on that day, obtained from supply data (source: ANAC Air Transport Statistics, own calculations), multiplied by 100. The variable indicates the company's competitive position in the route market.

- RHHI is the route concentration index, calculated based on the companies' seat share (source: ANAC Air Transport Statistics, own calculations). The variable shows the level of competition between the companies operating the route.
- LASTROW is a dummy variable indicating that the passenger's seat is in the last row, combining the seat reported in the EPL Survey with cabin maps (source: scanned boarding passes from the EPL Survey, Panrotas guides). The variable represents a less desirable position within the cabin.
- EMERGEXIT is a dummy indicator for seats located at the emergency exit, obtained by cross-referencing the seat declared in the EPL Survey with cabin maps (source: scanned boarding passes from the EPL Survey, Panrotas guides). The variable represents seats with more legroom and specific rules of use.
- COMFORT is a dummy indicator for airline seats with increased pitch, determined by the seat listed on the EPL Survey boarding pass and its position on the cabin map (source: scanned boarding passes from the EPL Survey, Panrotas guides). Expanded pitch seating is a position in the cabin where the spacing between rows is greater than the standard for the rest of the aircraft, offering additional comfort within economy class itself. These seats were identified on Gol's Boeing 737 aircraft and on some of TAM's (now Latam) Airbus A319, A320, and A321 models, based on an analysis of seat maps.
- COMFORT (placebo) is a dummy variable that identifies seats equivalent to those with increased pitch, but present on aircraft of the same model and from the same airline that did not offer this internal configuration (source: scanned boarding passes from the EPL Survey, Panrotas guides). This variable serves as a control to verify whether any effects attributed to onboard comfort, as measured by COMFORT, result from classification errors or structural characteristics of the layout that are not related to the differentiated product.
- MIDDLE is a dummy indicator of middle seats, determined from the seat reported in the EPL Survey and confirmed by cabin maps (source: scanned boarding passes from the EPL Survey, Panrotas guides). The variable represents a more laterally confined position.
- MIDDLE × ADV are interactions between the MIDDLE variable and advance booking ranges classified as 1w, 2w, 3w, and more than 3w, where w corresponds to weeks (source: scanned boarding passes from the EPL Survey, Panrotas guides; EPL Survey questionnaires, own calculations). The variables show how the effect of the middle seat changes depending on the time of purchase.
- IROWDENS is a row density index. It is the ratio between the total number of rows installed on the aircraft and the maximum number of rows documented internationally for the same model (source: Panrotas guide and Seatguru website[19]). The index represents a proxy for the structural compactness of the cabin, indicating greater intensity of space use, reduced comfort, and potential reduction in cost per seat when its values are higher. The index is normalized to 100.
- IPITCH is a seat pitch index. It is the ratio between the actual average spacing between aircraft rows, measured in inches, and the highest internationally documented pitch value for the same model (source: Panrotas Guide and Seatguru website[20]). The calculation uses a weighted average pitch based on the distribution of rows in the different sections of the cabin. The index reflects a proxy for the longitudinal comfort offered by the seat configuration, increasing when the space per passenger is greater. The index is normalized to 100.

To estimate the proposed econometric model, we used the Post-Double-Selection LASSO (PDS-LASSO) approach, which is suitable for phenomena that require the use of many potential controls. The main motivation for using this procedure is the risk of bias due to omitted variables associated with the complexity of airline ticket pricing. Simple situations illustrate this problem, such as differences in demand between more competitive and less sought-after times, specific variations on certain days that simultaneously influence passenger flow and fare levels, or individual characteristics related to the traveler's profile. For example, older passengers may plan further in advance, while younger passengers may purchase closer to the flight date, and individuals with higher incomes or high travel frequency may have different purchasing patterns than others. To reduce these potential biases, the estimation strategy adopts a broad set of candidate $c_l$ controls that are subjected to LASSO penalty and retained only when they show a statistically relevant relationship. Next, the regressors of interest are estimated by Ordinary Least Squares, adding the previously selected controls. The estimated standard errors were robust to pairing airports.

---

[19] Website seatguru.com, according to research conducted in August 2021.
[20] Idem.

Regarding the controls penalized by PDS-LASSO, dummies for search dates (130) and flight departure times (24) were used. In more in-depth high-dimensional specifications, controls related to the airports involved (55) and an extensive set of passenger profile controls (1761) from the EPL survey were also included. The latter combine age groups, income brackets, gender, and frequency of travel in the last year, which allows us to represent, for example, groups of younger or older passengers, those with lower or higher incomes, men, or women, and occasional or frequent travelers. Each combination generates a specific individual profile dummy that enters as a candidate in the penalty process, helping to absorb unobserved components of purchasing behavior that may simultaneously influence the price paid and the passenger's position in the cabin.

## VII. ESTIMATION RESULTS

The estimation results are presented in Table 3. The table uses eight columns to show how the estimates change as the regressor variables are introduced cumulatively, starting with the simplest specification on the left and moving to more complete specifications on the right. Column (5) corresponds to the central specification for interpretation. Columns (6) to (8) offer variations of this same structure, either by including interactions with MIDDLE in Column (6), incorporating high-dimensional controls for airports and passenger characteristics in Column (7), or replacing IROWDEN with the IPITCH proxy in Column (8).

ADV is negative and statistically significant in all specifications. This indicates an association between earlier purchase and lower fares, which is consistent with known patterns of revenue management algorithms. DIST is positive in all columns and remains statistically significant even after the inclusion of FLTIME, suggesting that distance and flight duration capture distinct components of unobservable operating costs. In Column (1), without FLTIME, DIST absorbs most of this variation. When FLTIME is included in the model from Column (3) to (8), the DIST coefficient decreases, indicating that both now share the statistical explanation of this cost component, with possible partial overlap between the two measures.

BSN is positive and statistically significant in Columns (2) to (8), with little variation between specifications, which signals stability of the estimated coefficient. This suggests that the fare difference for business passengers is not sensitive to the set of controls included in the different specifications and does not appear to result from systematic omission of variables in this sample. This behavior is consistent with airlines' fare differentiation practices, which, with segmentation schemes, can partially confer a price premium from business travelers and on routes more strongly marked by the predominance of these passengers in the consumer mix.

In the other proxies related to operational, strategic, and market aspects, the signs also remain relatively consistent across columns. FUELP is positive and statistically significant in all specifications, suggesting partial pass-through of fuel shocks. HUB has a stable positive coefficient, consistent with network characteristics and possible differences in market power at hub airports. RHHI is positive and statistically significant whenever included, with higher values in Columns (7) and (8), indicating that, with more detailed controls for passenger heterogeneity, the association between concentration and fare becomes more pronounced. This interpretation should be viewed with caution, given the possibility of omitted variables and identification limitations.

In the capacity and occupancy variables, it is observed that the coefficients follow the expected pattern but lose strength as the controls become more granular. LF is positive and statistically significant in Columns (3) to (6) and ceases to be statistically significant in Columns (7) and (8), indicating that, after controlling for passenger profiles and airports, the residual occupancy factor explains less fare variation. REVPAX is negative and statistically significant up to Column (6) and weakens thereafter, indicating that proxies for onboard traffic density capture fare composition in specifications where fixed effects are more numerous. SEATSH is also positive and statistically significant in Columns (3) to (6), losing significance in Columns (7) and (8), consistent with the fact that part of the information it carries is absorbed by airport effects and passenger characteristics. The same is true for SHIPMENT.

The interpretation of seat density and location variables (LASTROW, EMERGEXIT COMFORT, MIDDLE, IROWDENS, and IPITCH) must begin with the fact that, during the period analyzed, there was still no system for charging seat selection fees. This was a time when this practice was being introduced by the industry, especially in the front rows associated with extended pitch seats. Thus, any coefficient estimated for these variables reflects only differences in the price paid by passengers who occupied each type of seat, and not the effect of a specific fare. A statistically significant coefficient only indicates that, on average, those who occupied that seat paid a higher or lower amount, conditional on the model controls. This analysis allows us to understand the possibilities of charging selection fees for reordering passengers on board with a view to greater allocative efficiency. The results obtained suggest some interesting patterns, but the observed variability and limitations in the approach and estimator indicate that the topic still requires further investigation to better clarify the mechanisms involved.

IROWDENS is negative and statistically significant in Columns (1) to (6) and remains negative in Column (7), with a lower coefficient and statistical significance in this specification. This indicates that higher seat density is possibly associated with lower prices, which is consistent with the idea that increasing the number of rows reduces the cost per seat. These results suggest the presence of economies linked to the effective size of the cabin in the pricing process. In Column (8), we use another way to measure this same mechanism, IPITCH, which replaces IROWDENS and has a positive and statistically significant coefficient. In this case, greater average spacing between rows appears to be associated with higher prices, reflecting the same principle from another metric. The change in indicator maintains the economic interpretation and minimally alters the model's adjustment criteria. Because they are indicators constructed from external density or spacing references adopted by the industry in different operators and countries, measures such as IROWDENS and IPITCH reflect structural characteristics of the aircraft configuration, rather than direct passenger perceptions. Passengers only observe the space available in their own seats, without reference to the relative densities used by other airlines or manufacturers. Therefore, these indicators tend to capture unobservable components related to costs and the technical configuration of the cabin, rather than explicit consumer preferences. For this reason, the negative signs in IROWDENS (Columns 1 to 7) and the positive sign in IPITCH (Column 8) should not be interpreted as evidence of greater or lesser willingness to pay on the part of passengers about cabin layout. These coefficients mainly reflect structural variations in aircraft configuration and cost components associated with space usage, rather than differences directly perceived or valued by the user at the time of purchase.

The cabin location attributes measured by dummy variables (LASTROW, EMERGEXIT, COMFORT, and MIDDLE) show patterns that, in some cases, differ from what would be expected considering current practices. A statistically significant coefficient for these variables indicates that, on average, those who occupied that seat paid a higher or lower price, conditional on the model controls. However, LASTROW is not statistically significant, indicating no systematic difference in the price paid by passengers allocated to the last row, after controlling for other factors. EMERGEXIT is also not statistically significant, even though it is a position that today usually has a specific charge. This suggests that, during that period, this location did not generate detectable variations in the price paid by the occupants of these seats. COMFORT has positive coefficients, but is statistically significant only at the 10 percent level in Columns (5) to (8). This indicates limited evidence that passengers seated in seats with increased pitch paid slightly higher ticket prices. This interpretation is consistent with the presence of groups with a greater willingness to pay. The COMFORT placebo is not statistically significant, suggesting that when comparing aircraft of the same model and from the same airline that did not have seats with increased pitch, we are apparently not dealing with a spurious effect. This reduces the possibility that the coefficient observed in COMFORT results from measurement error or structural factors unrelated to the seat itself. However, the absence of significance at widely accepted statistical levels indicates that the COMFORT results do not clearly differ from the placebo, suggesting that, during this period, the offer of extra space did not play an effective segmentation role in the price paid by passengers.

MIDDLE presents a pattern that requires more detailed analysis. In Column (5), the estimated coefficient of this dummy is positive and statistically significant, a counterintuitive result, but conceptually possible in the context analyzed, which, as discussed, corresponds to a period prior to the charging of seat reservation fees. In this scenario, passengers who were more willing to pay for their tickets often arrived at the booking counter at short notice, finding that the side seats were already taken and that mostly center seats remained. This behavior reveals the economic incentives that airlines must introduce seat selection fees to strengthen the effectiveness of their segmentation schemes, to direct the best seats to passengers who are more willing to pay and more loyal, who tend to value the attributes of the service more. This provides evidence that apparently justifies the economic rationality of this type of charge in the industry.

The MIDDLE × ADV interactions, presented in Column (6) and the others, further clarify the mechanism that the data points to. The interactions indicate that passengers who pay more and end up sitting in the middle seat are possibly those who purchase very close to the flight date, especially within a week of the flight. For ADV of one week (1w), the estimated conditional effect is positive and statistically significant. For two weeks (2w), the effect is still positive, but smaller and statistically significant only at 10%. For 3 weeks (3w), the effect dissipates. For more than 3 weeks (>3w), the coefficient becomes negative and statistically significant only at the 10% level. In other words, the results provide evidence that those who arrived later to book found fewer seat options, and higher occupancy meant that mainly middle seats were left. The apparent ticket premium associated with this type of seat, therefore, does not reflect preference, but rather the dynamics of late purchasing in a period without seat reservation fees.

Adjustment criteria and specification choice indicate a preference for the most complete columns. AIC and RMSE decrease monotonically as more controls and cabin attributes are added. Columns (6) and (7) combine completeness with maintenance of IROWDENS and have low AIC and RMSE values, making them suitable when the objective is to directly measure the relationship between row density and prices. Column (8) has the lowest AIC and RMSE, at the cost of replacing the density metric with average spacing, functioning as a robustness test and as an interpretation in terms of willingness to pay.

In summary, three structural points stand out. First, cabin density matters for price formation, whether measured by IROWDENS or IPITCH, with opposite but consistent signals compatible with economies of scale associated with aircraft size measured in number of rows of seats. This result directly dialogues with the evidence from Lee & Luengo Prado (2004) that greater space per passenger can, in certain contexts, produce fare differences, although in our case the granularity of the microdata allows us to identify internal effects within the aircraft that their study did not observe. Second, the middle seat does not have a negative premium of its own, as would be expected, and its effect stems from the pattern of late purchases, as evidenced by interactions with ADV, indicating that, at the time, passengers with higher willingness-to-pay profiles often ended up in less desirable positions in the cabin. This finding broadens the interpretation of the literature by revealing that individual preferences can be masked by allocation mechanisms, something not captured by the design before and after structural changes in pitch analyzed by those authors. Third, variables linked to the competitive and network context, such as RHHI and HUB, raise prices, while aggregate indicators such as LF and REVPAX lose relevance when the model includes fine controls for date, airport, and profile, reinforcing that intra-aircraft heterogeneity adds a layer of variation that is not present in studies based on market averages. Finally, the results indicate that the sample period significantly conditions these relationships. As this is a phase prior to the systematic use of differentiation instruments associated with seat location, the patterns observed suggest limited effectiveness of the segmentation mechanisms that existed at the time. This points to room for improvement that would be explored in the following years, with the gradual adoption of practices aimed at distinguishing between products and allocating passengers more efficiently according to their preferences and willingness to pay.

**Table 3 – Results of estimates and robustness checks**

|  | (1) | (2) | (3) | (4) | (5) | (6) | (7) | (8) |
|---|---|---|---|---|---|---|---|---|
| ADV | -0.0044*** | -0.0031*** | -0.0032*** | -0.0032*** | -0.0033*** | -0.0029*** | -0.0030*** | -0.0030*** |
| DIST | 0.5320*** | 0.5516*** | 0.2626*** | 0.2362*** | 0.2385*** | 0.2390*** | 0.3151*** | 0.3090*** |
| BSN |  | 0.3993*** | 0.3941*** | 0.3868*** | 0.3912*** | 0.3827*** | 0.3828*** | 0.3829*** |
| FLTIME |  |  | 0.4401*** | 0.4824*** | 0.4776*** | 0.4790*** | 0.3018*** | 0.3056*** |
| SHIPMENT |  |  | 0.0382*** | 0.0356** | 0.0376*** | 0.0382*** | 0.0294* | 0.0220 |
| REVPAX |  |  | -0.2849*** | -0.1725*** | -0.1698*** | -0.1694*** | -0.1034* | -0.0893 |
| LF |  |  | 0.2648*** | 0.1766*** | 0.1632*** | 0.1575*** | 0.0722 | 0.0609 |
| FUELP |  |  | 1.0256*** | 0.7828*** | 0.7424*** | 0.7554*** | 0.8568*** | 0.7968*** |
| HUB |  |  | 0.5302*** | 0.5316*** | 0.5306*** | 0.5363*** | 0.5025*** | 0.5043*** |
| SEATSH |  |  |  | 0.0675*** | 0.0668*** | 0.0674*** | 0.0274 | 0.0235 |
| RHHI |  |  |  | 0.1457*** | 0.1490*** | 0.1477*** | 0.1887*** | 0.2011*** |
| LASTROW |  |  |  |  | 0.0322 | 0.0370 | 0.0503 | 0.0455 |
| EMERGEXIT |  |  |  |  | 0.0306 | 0.0268 | 0.0376 | 0.0455 |
| COMFORT |  |  |  |  | 0.6669* | 0.6602* | 0.6512* | 0.6600* |
| COMFORT (placebo) |  |  |  |  |  | 0.2207 | 0.1644 | 0.1816 |
| MIDDLE |  |  |  |  | 0.0649*** |  |  |  |
| MIDDLE × ADV (1w) |  |  |  |  |  | 0.2949*** | 0.2888*** | 0.2881*** |
| MIDDLE × ADV (2w) |  |  |  |  |  | 0.1018** | 0.0869* | 0.0854* |
| MIDDLE × ADV (3w) |  |  |  |  |  | -0.0232 | -0.0326 | -0.0347 |
| MIDDLE × ADV (>3w) |  |  |  |  |  | -0.0358* | -0.0367* | -0.0359* |
| IROWDENS | -0.6120*** | -0.5464*** | -0.6761*** | -0.7222*** | -0.6872*** | -0.7293*** | -0.4062** |  |
| IPITCH |  |  |  |  |  |  |  | 1.0998*** |
| Estimator | PDS/LASSO | PDS/LASSO | PDS/LASSO | PDS/LASSO | PDS/LASSO | PDS/LASSO | PDS/LASSO | PDS/LASSO |
| Airport-Pair Clusters | 333 | 333 | 333 | 333 | 333 | 333 | 333 | 333 |
| Svy Date Controls | 1/130 | 4/130 | 51/130 | 53/130 | 60/130 | 61/130 | 45/130 | 45/130 |
| Flight Time Controls | 3/24 | 3/24 | 4/24 | 4/24 | 4/24 | 4/24 | 2/24 | 2/24 |
| Airport Controls | No | No | No | No | No | No | 19/55 | 21/55 |
| PAX Profile Controls | No | No | No | No | No | No | 230/1761 | 230/1761 |
| AIC Statistic | 33,398 | 32,230 | 31,409 | 31,287 | 31,271 | 31,154 | 30,819 | 30,792 |
| BIC Statistic | 33,460 | 32,322 | 31,913 | 31,822 | 31,890 | 31,812 | 32,387 | 32,368 |
| Adj R2 Statistic | 0.2111 | 0.2681 | 0.2972 | 0.3029 | 0.3041 | 0.3096 | 0.3245 | 0.3256 |
| RMSE Statistic | 0.7039 | 0.6780 | 0.6643 | 0.6616 | 0.6611 | 0.6585 | 0.6513 | 0.6508 |
| Nr Observations | 15,634 | 15,634 | 15,517 | 15,517 | 15,517 | 15,517 | 15,517 | 15,517 |

*Notes: Estimation results in columns (1)-(8) produced by the LASSO-based methodology after double selection by Belloni et al. (2012, 2014a,b) (PDS-LASSO). Standard errors robust to heteroscedasticity. Estimates of omitted control variables. All control variables are penalized by LASSO. Representations of the P value: \*p<0.01, p<0.05, \* p<0.10.*

# VIII. CONCLUSIONS

This study investigated the relationship between cabin layout and airfare prices paid by travelers on domestic flights, based on data from a questionnaire survey conducted at Brazilian airports in the mid-2010s. The objective was to investigate whether physical elements of the internal configuration of aircraft and density indicators are associated with price variations, controlling for operational, competitive, network, and passenger profile factors. It is known that, currently, these onboard attributes are widely used by airlines to differentiate products and services, to grant benefits to premium customers and align the intrinsic quality of the seat with the passenger's willingness to pay. The more these dimensions of the firms' sales efforts are reconciled, the more efficient and competitive they become in the market. By examining this topic at a time when such practices were still in their infancy, it is possible to identify potential avenues for improvement and understand more precisely how seat allocation was being managed by companies and has evolved since then.

The estimation strategy employed the PDS-LASSO method, which selects variables in databases with many covariates to model complex phenomena. In the present study, the set of controls considered included effects of survey date, flight time, origin and destination airports, and passenger profile. These blocks were subjected to the selection procedure, which extracts relevant subsets for each specification, considering the statistical relationship of these covariates with price and variables of interest.

Some of the main findings were as follows. There is evidence of a negative association between seat density and prices, suggesting that aircraft with a higher number of rows per cabin unit tend to have lower fares, consistent with cost dilution effects. This point is relevant in the context of the Skyrider seat proposal (manufacturer Aviointeriors), discussed throughout the paper, which is configured as a concept of seats in an almost vertical position, in the shape of a bicycle saddle, allowing for substantial increases in the number of passengers on board. Other concepts can also produce similar effects of significantly increasing density, such as the Air Lair with two-story pods (Factorydesign), the Zephyr Seat with an additional upper level (Zephyr Aerospace), and the Cabin Hexagon with optimized geometry for adding seats (Zodiac Seats). The results obtained in this study indicate that increases in density are statistically associated with lower prices, suggesting that innovations in layouts that increase capacity, such as these and other seat and cabin layout designs currently under development, could benefit airlines, particularly those focused on attracting passengers with high price elasticity of demand. However, as these indicators capture structural effects rather than revealed passenger preferences, the economic viability of proposals such as Skyrider depends on how much consumers would be willing to accept reductions in comfort in exchange for lower fares, something that the data from this period does not allow us to assess directly.

Additionally, it was observed that middle seat occupants paid, ceteris paribus, higher ticket prices than aisle or window passengers, a pattern that interactions prior to purchase indicated was associated with late purchase dynamics, rather than any specific appreciation of this type of seat. This evidence is consistent with the fact that, in commercial environments without explicit seat selection fees, more desirable positions on board tend not to be as available when prices are already high due to revenue management mechanisms. This allocative inefficiency helps explain why seat selection pricing becomes an important market segmentation tool, as it allows companies to organize seat distribution in a manner more aligned with passengers' willingness to pay. The study therefore contributes by showing an economic basis that helps explain the rationale behind one of the main sources of ancillary revenue for airlines.

Finally, other location variables, such as extra-legroom seats, emergency exit seats, and last-row seats, were not statistically significant, indicating potential for investment in more effective segmentation practices. All these results are consistent with strategies that airlines have begun to develop intensively, in which customer segmentation strategies have become one of the most striking advances in recent commercial aviation. The growing use of onboard attributes, including dynamic pricing for seat selection and improvements in comfort differentiation, points to the continuity of this process and suggests that similar practices may continue to expand in the coming years, keeping pace with technological and commercial developments in cabin layouts.

## ACKNOWLEDGMENTS

The first author would like to thank the National Council for Scientific and Technological Development (CNPq), grant no. 305439/2021-9, and the São Paulo Research Foundation (FAPESP), grant no. 2024/01616-0. The authors would like to thank Cicero Rodrigues de Melo Filho, Daniel Klinger Vianna, Denise Deckers do Amaral, Matheus Lemos de Andrade, Thaynna Bocos, Mauro Caetano, Marcelo Guterres, Evandro Silva, Giovanna Ronzani, Rogéria Arantes, Cláudio Jorge Alves, and Mayara Murça. All errors are attributed to the authors.